\documentclass[reprint,twocolumn,notitlepage,secnumarabic,amssymb, amsmath, aps,nofootinbib, superscriptaddress]{revtex4-1}%\documentclass[letterpaper]{article}
\pdfoutput=1

\usepackage{amsmath}
\usepackage{amsfonts}
\usepackage{amsthm}
\usepackage{amssymb}
\usepackage{latexsym, array,multirow,  verbatim, enumerate}
\usepackage{slashed}
\usepackage{booktabs}
\usepackage{tabularx}
\usepackage{cancel}
\usepackage{dcolumn}
\usepackage{bm}
\usepackage{graphicx, caption, subcaption}  
\usepackage{url}
\usepackage[colorlinks]{hyperref}
\usepackage{color}
\usepackage[normalem]{ulem}
\usepackage{enumitem}
\usepackage{bbding}
\usepackage[compat=1.0.0]{tikz-feynman}
\usepackage{pifont}
\usepackage{empheq}
\usepackage[utf8]{inputenc}
\usepackage{bm}
\usepackage{blkarray}
\usepackage[font=footnotesize,labelfont=sf]{caption}
\usepackage{cleveref}

\def\nn{\nonumber}

\begin{document}
\title{Astrophysical hints for magnetic black holes}

\author{Diptimoy Ghosh, }
 \email{diptimoy.ghosh@iiserpune.ac.in}
\author{Arun Thalapillil, }
 \email{thalapillil@iiserpune.ac.in}
\author{and Farman Ullah}
 \email{farman.ullah@students.iiserpune.ac.in}
\affiliation{Department of Physics, Indian Institute of Science Education and Research Pune, Pune 411008, India}

\date{\today}
%%%%%%%%%%%%%%%%%%%%% Abstract %%%%%%%%%%%%%%%%%%%%%%%
\begin{abstract}
We discuss a cornucopia of potential astrophysical signatures and constraints on magnetically charged black holes of various masses. As recently highlighted, being potentially viable astrophysical candidates with immense electromagnetic fields, they may be ideal windows to fundamental physics, electroweak symmetry restoration and non-perturbative quantum field theoretic phenomena. We investigate various potential astrophysical pointers and bounds---including limits on charges, location of stable orbits and horizons in asymptotically flat and asymptotically de Sitter backgrounds, bounds from galactic magnetic fields and dark matter measurements, characteristic electromagnetic fluxes and tell-tale gravitational wave emissions during binary inspirals. Stable orbits around these objects hold an imprint of their nature and in the asymptotically de Sitter case, there is also a qualitatively new feature with the emergence of a stable outer orbit. We consider binary inspirals of both magnetic and neutral, and magnetic and magnetic, black hole pairs. The electromagnetic emissions and the gravitational waveform evolution, along with inter-black hole separation, display distinct features. Many of the astrophysical signatures may be observationally glaring---for instance, even in regions of parameter space where no electroweak corona forms, owing to magnetic fields that are still many orders of magnitude larger than even Magnetars, their consequent electromagnetic emissions will be spectacular during binary inspirals. While adding new results, our discussions also complement works in similar contexts, that have appeared recently in the literature.
\end{abstract}

\maketitle
%%%%%%%%%%%%%%%%%% Section %%%%%%%%%%%%%%%%%%%%%%%%%%%
\section{Introduction}{\label{sec:intro}
In our pursuit to understand the fundamental interactions that govern our universe, it is crucial to explore diverse phenomena that may inform us. One such avenue is provided by astrophysical objects with extreme properties. They are potentially spectacular probes for fundamental physics and beyond Standard Model phenomena, that are many times beyond the reach of terrestrial experiments. Rather than being merely relegated to fantasy, these avenues are coming to fruition with the discovery of gravitational waves from black hole/neutron star mergers\,\cite{Abbott:2016blz, TheLIGOScientific:2017qsa}, observation of near horizon features of black holes\,\cite{Akiyama:2019cqa}, discovery of evermore exotic astrophysical objects\,\cite{Mereghetti:2008je,Lorimer:2007qn,Thornton:2013iua}, and the advent of a multi-messenger era in astronomy\,\cite{GBM:2017lvd}.

One such class of possible exotic astrophysical objects are magnetically charged black holes (MBHs). Especially in the extremal or near-extremal limit, they may be relatively long lived cosmologically and potential survivors from our universe's early epochs. Furthermore, the paucity of magnetic monopoles and magnetically charged matter contribute to their persistence, making them amenable to current observations, if they have survived from some primordial cosmological era. 

MBHs and some of their intriguing theoretical aspects have been discussed in the past\,\cite{Lee:1991vy,Lee:1991qs,Lee:1994sk}. Recently, various spectacular properties of MBHs have been highlighted\,\cite{Maldacena:2020skw}, drawing on results from\,\cite{Ambjorn:1988tm,Ambjorn:1989sz,Ambjorn:1989bd,Ambjorn:1992ca}. As extensively discussed and emphasised in\,\cite{Maldacena:2020skw}, such objects may be windows to the nature of fundamental symmetries and other fundamental physics, hitherto inaccessible to other probes.  The magnetic fields near the horizon are generically large for these objects. In certain regions of
parameter space the magnetic fields may in fact even be large enough to restore electroweak symmetry\,\cite{Maldacena:2020skw,Ambjorn:1988tm,Ambjorn:1989sz,Ambjorn:1989bd,Ambjorn:1992ca}. Binary pairs of such objects, with opposite magnetic charges and in the extremal limit, also allow us to speculate on many intrguing possibilities\,\cite{Maldacena:2020sxe}. Thus, finding astrophysical hints for such objects, and subsequent observations of them, may have great repercussions for our understanding of the universe.

There have been some  phenomenological studies on MBHs recently\,\cite{Bai:2020spd, Liu:2020vsy}. In the low-mass MBH region, where an electroweak corona forms, a comprehensive phenomenological study has been performed in\,\cite{Bai:2020spd}. The effective one-body motion of a dyonic black hole binary system, along with some aspects of the corresponding electromagnetic and gravitational radiation, incorporating  post-Newtonian corrections, was studied recently in\,\cite{Liu:2020vsy}. Our study will be complementary to these, with a slightly different focus. There have also been studies in other contexts in the literature---for instance, studying implications of a topologically induced black hole electric charge\,\cite{Kim:2020bhg}, and interesting studies on MBH solutions arising in non-linear electrodynamics, and their implications on black hole shadows\,\cite{Allahyari:2019jqz}. In a broader context, with recent advances\,\cite{Abbott:2016blz, TheLIGOScientific:2017qsa, Akiyama:2019cqa}, it is also of great interest to probe and understand the exact nature of possible compact objects in our universe\,\cite{Cardoso:2019rvt}. For instance, horizon-less exotic compact objects may have a small reflectivity, with consequent observable manifestations in the post-merger ringdown and echoes\,\cite{Maggio:2020jml}. Compact objects carrying magnetic charges may introduce new elements to these considerations\,\cite{Clarkson:2003mp, Sotani:2013iha}. Quasinormal modes\,\cite{Vishveshwara:1970zz,Press:1971wr} for charged and uncharged compact objects\,\cite{Mellor:1989ac, Andersson:1996xw,Kokkotas:1999bd,Natario:2004jd}, in diverse spacetimes, have been studied previously in the literature. It would be interesting to study their astrophysical implications further, as a probe of fundamental physics and to deduce the exact characteristics of compact objects\,\cite{Chirenti:2017mwe, Bhagwat:2019dtm}.

 Our aim in this work is to investigate and discuss various potential astrophysical signatures for MBHs, while giving simple analytic results. As alluded to already, in the extremal and near-extremal limit (or the analogous limit in a de Sitter background) MBHs are expected to be long-lived and astrophysically viable. Moreover, as we shall see, across diverse MBH masses and magnetic charges, the magnetic fields near the horizon and near the innermost stable orbits can be extremely large. For instance, even for MBH masses where no electroweak corona\,\cite{Maldacena:2020skw} forms, these electromagnetic fields are many orders of magnitude bigger than even some of the currently known largest astrophysical magnetic fields---that of neutron stars such as Magnetars\,\cite{Mereghetti:2008je}. These unique characteristics, among others, may provide interesting astrophysical signatures for these objects.
 
 We will discuss simple bounds on the electric and magnetic charges for black holes from astrophysical considerations, features of astrophysically relevant stable circular orbits around MBHs in asymptotically flat and de Sitter backgrounds, limits from galactic magnetic field observations on MBH abundance when they are considered as a dark matter component, and characteristics of their electromagnetic and gravitational radiation, along with orbital evolution, during MBH binary inspirals. Our focus will not be just confined to parameter regions where an electroweak corona forms, but on relatively the full parameter space of MBH masses; in the extremal and near extremal limits. Most of the results and estimates we present are new. Where there are minor overlaps, our results are consistent, within astrophysical uncertainties and modelling assumptions, with the extant studies.
 
 In Sec.\,\ref{sec:bhsol} we briefly review the Reissner-Nordstrom solution, in asymptotically flat and de Sitter spacetimes. Here, we also comment on various limits on the electromagnetic charges from astrophysical considerations and discuss the location and relevant properties of the various horizons.
In Sec.\,\ref{sec:iscos} we then investigate, in some detail, stable circular orbits in the vicinity of extremal MBHs, considering both a vanishing and a positive cosmological constant background. In Sec.\,\ref{sec:ewsr} we briefly review certain quantum field theoretic aspects of extremal and near-extremal MBHs, with various $Q_B$. In this section, over this $Q_B$ range, we also present bounds on MBH as a dark mater candidate from galactic magnetic field and dark matter density measurements. In Sec.\,\ref{sec:ememissions} and Sec.\,\ref{sec:gwemissions} we then consider electromagnetic and gravitational wave emission from binary inspirals where MBHs are involved. We summarise our results and conclude in Sec.\,\ref{sec:summary}.

 %%%%%%%%%%%%%%%%%% Section %%%%%%%%%%%%%%%%%%%%%%%%%%%
\section{Astrophysical magnetically charged Reissner-Nordstrom black holes}
\label{sec:bhsol} 

In asymptotically flat spacetime, the standard Reissner-Nordstrom exterior solution of a black hole carrying electric and magnetic charges is of the form
\begin{eqnarray}
ds^2 &=& -\left(1-\frac{2M}{r}+\frac{\left(Q_E^2+Q_B^2\right)}{r^2}\right) dt^2\nn \\
&+& \left(1-\frac{2M}{r}+\frac{\left(Q_E^2+Q_B^2\right)}{r^2}\right)^{-1} dr^2+r^2 d\Omega^2 \; .
\end{eqnarray}
We are using units where $c=1, G_N=1, 1/4 \pi \epsilon_0 = \mu_0/4 \pi=1$.
$(r,\theta,\phi)$ are the coordinates, and $M$ is the mass of the black hole. $Q_E$ and $Q_B$ are the electric and magnetic charges of the black hole respectively. The dimensionless magnetic charge, $Q$, as defined in\,\cite{Maldacena:2020skw}, is related to $Q_B$ in our convention through $Q=Q_B/q_B$, where $q_B = 2 \pi \epsilon_0 \hbar c^2/e = 3.44 \times 10^{-9}$ A-m; it is the basic unit of magnetic charge assuming Dirac quantization.

In most realistic astrophysical and cosmological settings, it is relatively easy for the electric charge ($Q_E$) to get neutralised by in-falling matter or be strongly limited by other physical considerations. Consider for instance an electrically charged Reissner-Nordstrom black hole of mass $M$ and electric charge $+Q_E$. For an ionised Hydrogen nuclei to be absorbed, negating electrostatic repulsion, so as to increase its intrinsic charge, one requires conservatively
\begin{equation}
\frac{Q_E}{M}\lesssim \frac{m_p}{|e|}\simeq 10^{-18} \; .
\label{eq:qeattarg}
\end{equation}
Here, $m_p$ is the proton mass. This implies that in realistic situations the ratio will be very small in general. This assumes that the ions are not being impinged on the black hole with a large kinetic energy. Imparting higher kinetic energy will weaken the bound slightly, but the general import of the result remains.

Another theoretical consideration is that when the black hole is electrically charged, the near horizon electric fields may source Schwinger pair production of particle anti-particle pairs\,\cite{1982PhRvD..25.2509H}, say of mass $m$ and charge $q$. For the same charge, the non-perturbative rate will be dominated by the lightest charged particle in the spectrum. The oppositely charged particle falls into the black hole, reducing its overall charge, while the same-charge particle is repelled outwards. As the black hole discharges, the near horizon field subsequently decreases below the critical value and pair production essentially stops. As the non-perturbative rate\,\cite{Schwinger:1951nm} goes as $\Gamma_E \sim \exp[-m^2/qE]$, it implies that when
\begin{equation}
\frac{Q_E}{(2 M)^2} \ll \frac{m^2}{q} \; ,
\end{equation}
subsequent discharge would be suppressed. This then places a limit on the equilibrium black hole electric charge
\begin{equation}
\frac{Q_E}{M} \lesssim 10^{-5} \left( \frac{e}{q}\right)  \left( \frac{m}{m_e}\right)^2  \left( \frac{M}{M_\odot}\right) \; .
\end{equation}
Again, one observes that generally the black hole charge to mass ratio is expected to be small, and well below the $Q_E=M$ extremal limit.

Even when an electrically charged Reissner-Nordstrom black hole potentially forms, it may also get quickly neutralised by accreting ionised plasma. Considering standard Eddington accretion rates\, \cite{Carroll:1009754} $\dot{M}_{\text{\tiny{Ed.}}}=4 \pi cG M m_p/\sigma_{\text{\tiny{Thom.}}} \sim 10^{-15} (M/M_\odot) M_\odot \, \mathrm{s}^{-1}$, for a black hole charge $Q_E$, the neutralisation timescale will be roughly
\begin{equation}
t_{\text{\tiny{neutr.}}} \sim \frac{Q_E}{q} \frac{m}{\dot{M}_{\text{\tiny{Ed.}}}}\simeq 10^{15} \left( \frac{Q_E}{q}\right) \left( \frac{m}{M}\right)~\mathrm{s} \; .
\end{equation}
Even in the extremal case $Q_E=M$, assuming neutralization by accretion of ionised hydrogen, the charge neutralization is very rapid with
\begin{equation}
t_{\text{\tiny{neutr.}}}^{Q_E=M} \sim 10^{-3} s \; .
\end{equation}
It has been commented that in certain very special situations a background magnetic field may be able to thwart neutralisation by charge-selective accretion\,\cite{1982PhRvD..25.2509H} and extend the effective time-scale. Other accretion models and rates may change this estimate as well, but the time-scales are still expected to be relatively short. The broad message therefore seems to be that electrically charged black holes, even when they form, may not be very astrophysically long-lived.

Note that unlike the electrically charged counterpart, a classical magnetically charged black hole may persist for much longer, owing to the paucity of magnetically charged matter and magnetic monopoles, or if the monopoles are very heavy. Considerations from both Schwinger pair production of magnetic monopoles and accretion of magnetic monopoles lead to much weaker constraints.

For instance, the rate of Schwinger pair production of magnetic monopoles in a homogeneous magnetic field\,\cite{Affleck:1981bma,Affleck:1981ag} goes as $\Gamma_B \sim \exp[-m_B^2/q_B B]$. The absence of monopoles might make this channel impossible, or even if they exist, the generic heaviness of viable magnetic monopoles would for all practical purposes stall the discharge.

Next, consider an MBH of charge $+Q_B$ and like-charged magnetic monopoles of charge $q_B$ and mass $m_B\sim 10^{16}\,\mathrm{GeV}$ participating in the charging process, through accretion. Repeating the arguments leading up to Eq.\,(\ref{eq:qeattarg}), one now obtains a much weaker bound
\begin{equation}
\frac{Q_B}{M}\lesssim \frac{m_B}{|q_B|}\simeq 10^{-4} \; .
\label{eq:qbattarg}
\end{equation}
Here, we have assumed Dirac quantization. 

We therefore see that the bounds on the charge-to-mass ratio for MBHs are many orders of magnitude weaker than that for electrically charged black holes. This then implies that MBHs may have larger charge-to-mass ratio relative to their electrically charged counterparts. Nevertheless, this also indicates that to achieve an even higher ratio ($\gtrsim 10^{-4}$), the candidate MBH's origin may have to be distinct from conventional astrophysical scenarios, like that of a low charge-to-mass ratio MBH just slowly capturing low-energy, like-charged magnetic monopoles. Of course, again, if sufficient kinetic energy may be imparted to the in-falling monopoles in some way, the bound may be weakened and some further charging is possible through this mode. A possibility is also that the only viable monopoles have small effective fractional charges\,\cite{Hook:2017vyc,Terning:2018lsv,Chandra:2019dnf}, but are also heavy, which may weaken the bound. Another more plausible origin is that the MBH forms far from extremality, but due to the enhanced Hawking radiation\,\cite{Maldacena:2020skw} it quickly tends to $\frac{Q_B}{M}\rightarrow 1$. This scenario is also more pertinent, since primordial cosmological epochs may have furnished an amenable avenue for MBHs with large charge-to-mass ratios to form\,\cite{Stojkovic:2004hz,Bai:2019zcd}. We will adopt an agnostic viewpoint about the exact details of their cosmological origin.

In keeping with the main theme of the paper, and due to the above points, we will assume $Q_E=0$ throughout.  Let us consider therefore the Reissner-Nordstrom solution purely with $Q_B\neq0$. The horizons for such a black hole are located at
\begin{equation}
r_\pm = M \pm \sqrt{M^2-Q_B^2}
\end{equation}
Note that when $M=|Q_B|$, the two horizons merge and one has an event horizon located at $r_H=|Q_B|$. This case will be referred to as the extremal MBH. The $r$ coordinate in this case is spacelike on either side of the horizon. The singularity at $r=0$ is timelike. We note for future reference that in physical units, the extremal MBH has $M = \sqrt{ \dfrac{\mu_0}{4 \pi G_N}} \, Q_B$. When $M<|Q_B|$ one obtains a naked singularity.

The extremal solutions have many intriguing properties. For instance, two well separated extremal black holes with like charges will repel each other, exactly cancelling their gravitational attraction. In contrast, two extremal black holes with unlike charges will attract each other, with a magnitude exactly matching their gravitational attraction. These features will be particularly relevant to our later discussions. Another important observation is that extremal magnetic black holes in asymptotically flat spacetime are cold ($T^{M=|Q_B|}=0$), and may hence be long lived due to the suppression of Hawking radiation. This opens the possibility of such extremal MBHs surviving from earlier epochs of the universe to the present day. This makes them fascinating candidates to constrain and search for in astronomical observations. 

Now consider the case when the cosmological constant is non-vanishing and positive ($\Lambda>0$). For a magnetic Reissner-Nordstrom black hole embedded in asymptotically de Sitter spacetime, the metric is given by
\begin{equation}
ds^2=-\Delta(r) dt^2 + \Delta(r)^{-1} dr^2+r^2 d\Omega^2 \; ,
\end{equation}
with
\begin{equation}
\Delta(r)= \left(1-\frac{2M}{r}+\frac{Q_B^2}{r^2}-\frac{\Lambda  r^2}{3}\right) \; .
\end{equation}

When $Q_B=0$ and $\Lambda < 1/(9 M^2)$, the horizons are located, as is well-known, at
\begin{eqnarray}
r_{+} &=& \frac{2}{\sqrt{\Lambda}} \cos\left[ \frac{1}{3} \arccos\left(-3 M \sqrt{\Lambda}\right) \right] \; , \nn \\
r_{-} &=& \frac{2}{\sqrt{\Lambda}} \cos\left[\frac{4\pi}{3}+ \frac{1}{3} \arccos\left( -3 M \sqrt{\Lambda}\right) \right] \; .
\end{eqnarray}
When $\Lambda = 1/(9 M^2)$ and $Q_B=0$, there is a single horizon at
\begin{equation}
r_{0}= \frac{1}{\sqrt{\Lambda}}=3 M \;.
\end{equation}
For $\Lambda > 1/(9 M^2)$, there are no horizons, and one has a naked singularity. Note also that for the $\Lambda < 1/9 M^2$ case,  $2M<r_{-}<3 M<r_{+}$. 

Assuming the small observational value of $\Lambda$, one can in fact rewrite the above expressions more simply as 
\begin{eqnarray}
r_{+} &=& \sqrt{\frac{3}{\Lambda}} \left( 1 - \frac{\epsilon}{4} + \mathcal{O} (\epsilon^2) \right),\nn \\ 
r_{-} &=& 2 M \left( 1 + \frac{\epsilon^2}{4} + \mathcal{O} (\epsilon^4) \right)\; ,
\end{eqnarray}
where, 
\begin{equation}
\epsilon = 4 M \sqrt{\frac{\Lambda}{3}} = 3.6 \times 10^{-23} \frac{M}{M_\odot} \; .
\end{equation}

For magnetically charged black holes in de Sitter space (MBHdS), and specifically for magnetic black holes with $M=|Q_B|$, the situation is much richer. When $M=|Q_B|$, the horizons are now located at
\begin{eqnarray}
r_{i} &=& \frac{\sqrt{3+4 \sqrt{3} \sqrt{\Lambda } |Q_B|}-\sqrt{3}}{2 \sqrt{\Lambda }} \; , \nn \\
r_{o}&=&\frac{\sqrt{3}-\sqrt{3-4 \sqrt{3} \sqrt{\Lambda } |Q_B|}}{2 \sqrt{\Lambda }} \; ,\nn \\
r_{c} &=& \frac{\sqrt{3}+\sqrt{3-4 \sqrt{3} \sqrt{\Lambda } |Q_B|}}{2 \sqrt{\Lambda }} \; ,
\end{eqnarray}
when $\Lambda < 3/(16 Q_B^2)$. These correspond to the inner Cauchy horizon, outer event horizon and the de Sitter cosmological horizon respectively; with $r_i<r_o<r_c$.

Again, for the observational value of $\Lambda$, we may rewrite these simply as
\begin{eqnarray}
r_{i} &&= \sqrt{\frac{3}{\Lambda}} \left\{ \frac{1}{2} \left(\sqrt{1+\epsilon} - 1\right) \right\}  = Q_B \left(1 - \frac{\epsilon}{4} + \mathcal{O} (\epsilon^2) \right) \; ,\nn \\
r_{o}&&= \sqrt{\frac{3}{\Lambda}} \left\{ \frac{1}{2} \left(1 - \sqrt{1-\epsilon} \, \right) \right\} = Q_B \left(1 + \frac{\epsilon}{4} + \mathcal{O} (\epsilon^2) \right)\; , \nn \\
r_{c} &&= \sqrt{\frac{3}{\Lambda}} \left\{ \frac{1}{2} \left(1 + \sqrt{1-\epsilon} \, \right) \right\} = \sqrt{\frac{3}{\Lambda}}  \left(1 - \frac{\epsilon}{4} + \mathcal{O} (\epsilon^2) \right)\; .~~~~~~
\end{eqnarray}

We note that due to the smallness of the observed cosmological constant, the inner and outer horizon separation will be very small in reality
\begin{equation}
r_o -r_i \simeq  \frac{\epsilon}{2}Q_B \; .
\end{equation}

One interesting point to note is that when $\Lambda>0$ and for $M=|Q_B|$ case, the outer horizon located at $r_{o}$ has the same temperature as the de Sitter cosmological horizon located at $r_{c} $. This common temperature is given by\,\cite{Romans:1991nq} (see appendix \ref{app-A})
\begin{equation}
T^{M=|Q_B|}_{\text{\tiny{MBHdS}}}=\frac{1}{2 \pi}\sqrt{\frac{\Lambda}{3}\left(1-4 M \sqrt{\frac{\Lambda}{3}}\right)} \simeq \frac{1}{2 \pi}\sqrt{\frac{\Lambda}{3}}\left( 1-\frac{\epsilon}{2}\right)\; .
\label{eq:rndseqtemp}
\end{equation}

Thus, assuming other black holes are sufficiently far away with distinct de Sitter horizons of their own, the $M=|Q_B|$ case may again approximate a thermodynamically stable state\,\cite{Romans:1991nq,Kastor:1992nn}. Then, such magnetic black holes embedded in an asymptotically de Sitter background may also ideally be relatively long lived and amenable to present-day astrophysical observations.

The horizons of an MBH provide a causal structure, while also defining characteristic length scales for various astrophysical phenomena near the singularity. Another set of length scales are provided by  stable orbits near the compact object, which we explore next. In the next section we will then also compare the location of the stable orbits relative to the location of the MBH horizons.

 %%%%%%%%%%%%%%%%%% Section %%%%%%%%%%%%%%%%%%%%%%%%%%%
%\newpage
\section{Stable circular orbits near MBHs}{
\label{sec:iscos}
%%%%%%%%%%%%%%%%%%%%%
In General Relativity, the innermost stable circular orbit (ISCO) of a compact object, as the name suggests, is the smallest stable circular orbit for a massive or massless test particle. The ISCOs are important in astrophysics---for instance, in black hole accretion disks, where they potentially mark a characteristic inner edge of the disk. During binary inspirals they also mark the boundary after which the compact objects begin the merge phase and are plunging into each other. The stable orbits provide information about the type and characteristics of the compact object.  Their study is therefore pertinent and even more prescient in the era where there are ongoing endeavours to observe near horizon features of black holes\,\cite{Akiyama:2019cqa}.

There have been extensive theoretical studies on stable orbits for test particles, in asymptotically flat Reisnner-Nordstrom spacetimes\,\cite{Kim:2007ca, Pradhan:2010ws,Grunau:2010gd,Pugliese:2011py,Pugliese:2013xfa,Gonzalez:2017kxt,Russo:2020lah}, as well as some studies in Reisnner-Nordstrom spacetimes with a non-zero cosmological constant\,\cite{1983BAICz..34..129S,Stuchlik:2002tj}. There has also been extensive studies on photon surfaces in arbitrary space-times\,\cite{Claudel:2000yi}. We would specifically like to investigate, and obtain simple analytic expressions for, stable orbits for MBHs having $M=|Q_B|$, in asymptotically flat and asymptotically de Sitter backgrounds, with a focus on potential astrophysical observables. For instance, one of the intriguing results that we will obtain is the observation that for MBHs in asymptotically de Sitter backgrounds there is a new qualitative feature (absent both when $\Lambda=0$ or $\Lambda<0$)---the emergence of outer stable circular orbits, that depend on $Q_B$. We will also glean a simple analytic expression for this boundary when $M=|Q_B|$. Similar observations have been made for Schwarzschild de Sitter spacetimes\,\cite{1999PhRvD..60d4006S, Boonserm:2019nqq} in the past. We also relate the location of various length scales, including those of the ISCOs, to comment on other observables in later sections.

To clarify concepts and for comparisons, let us briefly review the status of ISCOs for a Schwarzschild black hole (BH) in an asymptotically flat background. Considering $\theta=\mathrm{const.}$, with a time-like or null affine parametrization, one has
\begin{eqnarray}
\vartheta^2 &\equiv& g_{\mu\nu}\frac{dx^\mu}{ds} \frac{dx^\nu}{ds}\nn \\
&=& -\left(1-\frac{2M}{r}\right) \left( \frac{dt}{ds}\right)^2+\left(1-\frac{2M}{r}\right)^{-1} \left( \frac{dr}{ds}\right)^2 \nn \\
&+& r^2 \left( \frac{d\phi}{ds}\right)^2\; .~~~
\end{eqnarray}

$\vartheta^2$ is vanishing for massless particles, like photons, and equal to $-1$ for massive particles and bodies. Utilising the Killing symmetries for the Schwarschild spacetime, corresponding to energy ($E$) and angular momentum ($l$), the above may be reduced to
\begin{equation}
\left(\frac{dr}{ds}\right)^2= E^2+ \left(1-\frac{2M}{r} \right) \left( \vartheta^2- \frac{l^2}{r^2}\right) \; .
\label{eq:scheq}
\end{equation}
From above, one may define an effective potential of the form
\begin{equation}
V_{\text{\tiny{BH,eff}}}(r; \vartheta^2)=-\frac{1}{2} \left(1-\frac{2M}{r} \right) \left( \vartheta^2- \frac{l^2}{r^2}\right) \; .
\label{eq:scheffpot}
\end{equation}
This may now be analysed to deduce orbits and their stabilities.

For photons, $\vartheta^2=0$, and we obtain the effective potential extremum by solving
\begin{equation}
V'_{\text{\tiny{BH}},\gamma}(r;0)=-\frac{l^2}{r^3}+\frac{3 l^2 M}{r^4}=0 \; ,
\end{equation}
leading to the solution $r=3M$. From the sign of
\begin{equation}
V''_{\text{\tiny{BH}},\gamma}(r;0)=\frac{3 l^2}{r^4}  -\frac{12 l^2  M}{r^5} \; ,
\end{equation}
 for this solution, we conclude that
\begin{equation}
r_{\text{\tiny{BH}},\gamma}=3 M \; ,
\label{eq:szps}
\end{equation}
is an unstable photon sphere.

For massive particles, $\vartheta^2=-1$. Now, the extrema of the effective potential are at
\begin{equation}
V'_{\text{\tiny{BH}},m}(r;-1)=\frac{M}{r^2}-\frac{l^2}{r^3}+\frac{3 l^2 M}{r^4}=0 \; ,
\label{eq:szmassiveicco}
\end{equation}
For finite and real-valued angular momenta ($l$), the above equation has solutions for $r$ in the range $(3 M, \infty)$. One must analyse the stability of these orbits considering
\begin{equation}
V''_{\text{\tiny{BH}},m}(r;-1)= -\frac{2  M}{r^3} +\frac{3 l^2}{r^4}-\frac{12 l^2  M}{r^5}\; ,
\end{equation}
to identify the ISCO. Substituting for $l^2$ in above, from Eq. (\ref{eq:szmassiveicco}), we identify that $V''_{\text{\tiny{eff}}}(r; -1)$ transitions from positive values to zero at
\begin{equation}
r_{\text{\tiny{BH}},m}^{\text{\tiny{ISCO}}}=6 M \; .
\label{eq:szmisco}
\end{equation}
This corresponds to the well-know ISCO for massive particles in the Schwarschild spacetime background. Below this radius, in the range $(3M, 6M)$, the circular orbits are unstable for massive bodies. We note from Eqs. (\ref{eq:szps}) and (\ref{eq:szmassiveicco}) that below $3M$, no circular orbits can exist for both photons and massive bodies.

Let us now turn our attention to MBHs in asymptotically flat spacetimes. Following similar arguments, as for the Schwarschild spacetime, one obtains using the Killing symmetries
\begin{equation}
\left(\frac{dr}{ds}\right)^2= E^2+ \left(1-\frac{2M}{r} +\frac{ Q_B^2}{r^2} \right) \left( \vartheta^2- \frac{l^2}{r^2}\right) \; ,
\label{eq:rneq}
\end{equation}
with the corresponding effective potential identified as
\begin{equation}
V_{\text{\tiny{MBH,eff}}}(r; \vartheta^2)=-\frac{1}{2}\left(1-\frac{2M}{r} +\frac{ Q_B^2}{r^2} \right)\left( \vartheta^2- \frac{l^2}{r^2}\right) \; .
\label{eq:rneffpot}
\end{equation}

For massless particles like the photon, the extrema may again be identified by solving
\begin{equation}
V'_{\text{\tiny{MBH}},\gamma}(r;0)=-\frac{l^2}{r^3}+\frac{3 l^2 M}{r^4}-\frac{2 l^2 Q_B^2}{r^5}=0 \; ,
\label{eq:mrnvp}
\end{equation}
and their stabilities analysed considering the sign of
\begin{equation}
V''_{\text{\tiny{MBH}},\gamma}(r;0)=\frac{3 l^2}{r^4}  -\frac{12 l^2  M}{r^5}+\frac{10 l^2  Q_B^2}{r^6}\; ,
\label{eq:mrnvpp}
\end{equation}
at the respective extrema.

We focus on the extremal MBH with $M=|Q_B|$, in an asymptotically flat spacetime background. In this limit, we may analytically investigate Eqs. (\ref{eq:mrnvp}) and (\ref{eq:mrnvpp}) for the possibility of stable circular orbits. We observe that there is an ostensibly stable circular orbit at the horizon 
\begin{equation}
r_{\text{\tiny{MBH}},\gamma}^{0}= |Q_B| \; ,
\label{eq:rnps1}
\end{equation}
and an unstable circular orbit at 
\begin{equation}
r_{\text{\tiny{MBH}},\gamma}^{1}=2 |Q_B| \; ,
\label{eq:rnps2}
\end{equation}
corresponding to the photon spheres. The event horizon is located at $r_H= |Q_B|$, and any perturbation of the test photon near $r_{\text{\tiny{MBH}},\gamma}^{0}$, towards the horizon, will cause it to plunge through it. Thus, it is stable only in one direction.

For massive neutral bodies, the analogous extrema are located at solutions to
\begin{equation}
V'_{\text{\tiny{MBH}},m}(r;-1)= \frac{M}{r^2}-\frac{(Q_B^2+l^2)}{r^3}+\frac{3 l^2 M}{r^4}-\frac{2 l^2 Q_B^2}{r^5}=0 \; ,
\end{equation}
with their stabilities determined by the sign of
\begin{equation}
V''_{\text{\tiny{MBH}},m}(r;-1)=-\frac{2 M}{r^3}+\frac{3( l^2+ Q_B^2)}{r^4}-\frac{12 l^2 M}{r^5}+\frac{10 l^2 Q_B^2}{r^6}\; .
\label{eq:rnvpp}
\end{equation}

Again, in the interesting extremal limit $M=|Q_B|$, we are able to proceed analytically and investigate the status of stable circular orbits. 

The viable extrema are now at
\begin{eqnarray}
r_{\text{\tiny{MBH}},m}^{0}&=& |Q_B| \; , \nn \\
r_{\text{\tiny{MBH}},m}^{\pm}&=& \frac{l^2\pm\sqrt{l^4-8 l^2  Q_B^2}}{2 |Q_B|} \; .
\end{eqnarray}
Analysing Eq.\,(\ref{eq:rnvpp}), one notes that there is an apparently stable circular orbit at $r_{\text{\tiny{MBH,m}}}^{0} \forall~ l^2 \geq 0$. Again, due to the presence of the event horizon at $r_H= |Q_B|$, any perturbation of the massive test particle near $r_{\text{\tiny{MBH}},m}^{0}$ may cause it to fall in through the horizon. The orbit at $r_{\text{\tiny{MBH}},m}^{-}  \forall~l^2  >  8  Q_B^2$ is unstable. There are also stable circular orbits for $r_{\text{\tiny{MBH}},m}^{+} \forall~l^2  \geq 8  Q_B^2$. The smallest radii at which a truly stable orbit exists, and therefore the location of the ISCO, is at
\begin{equation}
r_{\text{\tiny{MBH}},m}^{\text{\tiny{ISCO}}}= 4  |Q_B| \; .
\label{eq:rnisco}
\end{equation}
Note also that stable orbits given by real-valued $r_{\text{\tiny{MBH}},m}^+  \forall~l^2  \geq 8  Q_B^2$ extend all the way to $+\infty$. We will note later that in a de Sitter background there is a qualitative difference.

Let us now consider the case of MBHs in asymptotically de Sitter spacetimes with a positive cosmological constant ($\Lambda >0$), and analyse the status of stable circular orbits therein.

For comparisons, again consider first the Schwarzschild de Sitter  spacetime (BHdS)--- i.e. with. $Q_B=0$ and $\Lambda >0$. For massless particles there is again an unstable photon sphere at
\begin{equation}
r_{\text{\tiny{BHdS}},\gamma}= 3  M \; ,
\end{equation}
and for massive particles an ISCO at around
\begin{equation}
r_{\text{\tiny{BHdS}},m}^{\text{\tiny{ISCO}}} \simeq  6M \left[ 1+\frac{81} {4} \epsilon^2+\mathcal{O}(\epsilon^4)\right]\; .
\end{equation}
We have assumed again that $M^2 \Lambda \ll 1$, consistent with the observed value of the cosmological constant and mass ranges we consider.

Interestingly, as has been pointed out recently\,\cite{1999PhRvD..60d4006S,Boonserm:2019nqq}, there is now an astrophysically intriguing qualitative difference in the outer orbits that is absent in asymptotically flat ($\Lambda=0$) or anti de Sitter  ($\Lambda < 0$) backgrounds. Instead of the stable or quasi stable outer orbits extending all the way to $+\infty$, there is now a boundary beyond which $l^2$ is no longer positive and finite. This limit is denoted by an outer stable or quasi-stable circular orbit (OSCO). For Schwarzschild de Sitter spacetime this is located approximately at\,\cite{1999PhRvD..60d4006S,Boonserm:2019nqq}
\begin{equation}
r_{\text{\tiny{BHdS}},m}^{\text{\tiny{OSCO}}} \simeq \left(\frac{3M}{4\Lambda}\right)^{1/3} \; .
\end{equation}
For some systems this boundary may potentially be of astrophysical or cosmological significance. For example, for  typical galactic masses they are of the same scale as the intergalactic spacing, and for galaxy clusters they are of roughly the same size as the clusters themselves\,\cite{Boonserm:2019nqq}.

Let us now turn to the $Q_B \neq 0$ case, assuming a background with a positive cosmological constant (MBHdS). We now need to analyse the effective potential 
\begin{equation}
V_{\text{\tiny{MBHdS,eff}}}(r; \vartheta^2) =-\frac{1}{2}\left(1-\frac{2M}{r} +\frac{ Q_B^2}{r^2} -\frac{\Lambda  r^2}{3} \right)\left( \vartheta^2- \frac{l^2}{r^2}\right) \; .
\label{eq:rndseffpot}
\end{equation}

For massless particles ($\vartheta^2=0$), when $M=Q_B$, the photon sphere (stable in one direction) is at the same radius as in the $\Lambda=0$ case,
\begin{equation}
r_{\text{\tiny{MBHdS}},\gamma}^{0}= |Q_B| \; .
\label{eq:rndsps1}
\end{equation}
There is an unstable circular orbit at $r_{\text{\tiny{RNdS}},\gamma}^{1}=2 |Q_B|$, also as in the asymptotically flat case.

For massive neutral bodies, the analogous extrema are located at
\begin{eqnarray}
V'_{\text{\tiny{MBHdS}},m}(r;-1) &=& -\frac{\Lambda r}{3}+\frac{M}{r^2}-\frac{(Q_B^2+l^2)}{r^3}+\frac{3 l^2 M}{r^4}-\frac{2 l^2 Q_B^2}{r^5}\nn \\
&=& 0 \; ,
\label{eq:rndsvp}
\end{eqnarray}
with their stabilities determined by the sign of
\begin{equation}
V''_{\text{\tiny{MBHdS}},m}(r;-1)=-\frac{\Lambda }{3}-\frac{2 M}{r^3}+\frac{3( l^2+ Q_B^2)}{r^4}-\frac{12 l^2 M}{r^5}+\frac{10 l^2 Q_B^2}{r^6}\; .
\label{eq:rndsvpp}
\end{equation}

For $M=|Q_B|$---in which case the outer event horizon and de Sitter cosmological horizon have equal temperatures given by Eq. (\ref{eq:rndseqtemp}), and may therefore denote a thermodynamically stable situation---Eq.\,(\ref{eq:rndsvp}) implies that
\begin{equation}
l^2=-\frac{r^2 \left[ \frac{\Lambda}{3} r^4-Q_B(r-Q_B)\right]}{(r-Q_B)(r-2Q_B)}\;.
\label{eq:rndslsq}
\end{equation}
This may be substituted in Eq.\,(\ref{eq:rndsvpp}) and analysed for vanishing points. 

At $r=4|Q_B|$, $V''_{\text{\tiny{MBHdS}},m}(4 Q_B;-1) <0$, and below this radius it has a negative sign as well. Hence, we conclude that the ISCO must now lie slightly above this radius. For the observed positive cosmological constant, for which $Q_B^2 \Lambda \ll 1$ in the range of interest to us, the location of the ISCO may be deduced by Newton-Raphson iterations and is found to be approximately at 
\begin{equation}
r_{\text{\tiny{MBHdS}},m}^{\text{\tiny{ISCO}}}\simeq 4  |Q_B| \left[1+\frac{64}{9} \epsilon^2+\mathcal{O}(\epsilon^4)\right]\; .
\label{eq:rndsisco}
\end{equation}

There is again a qualitatively new feature for the magnetic Reissner Nordstrom black hole embedded in a de Sitter background--the presence of an outer boundary for stable orbits---as earlier also observed in the Schwarschild de Sitter case. This feature may be deduced and investigated by analysing Eq.\,(\ref{eq:rndslsq}), cognisant of the requirement that for physically allowed orbits $l^2$ must be positive and finite.

%%%%%%%%%%%%%%%%%%%%%%%%%%%%%%%%%%%%
\begin{figure}[h]
\centering
\begin{tabular}{c}
\includegraphics[scale=0.3]{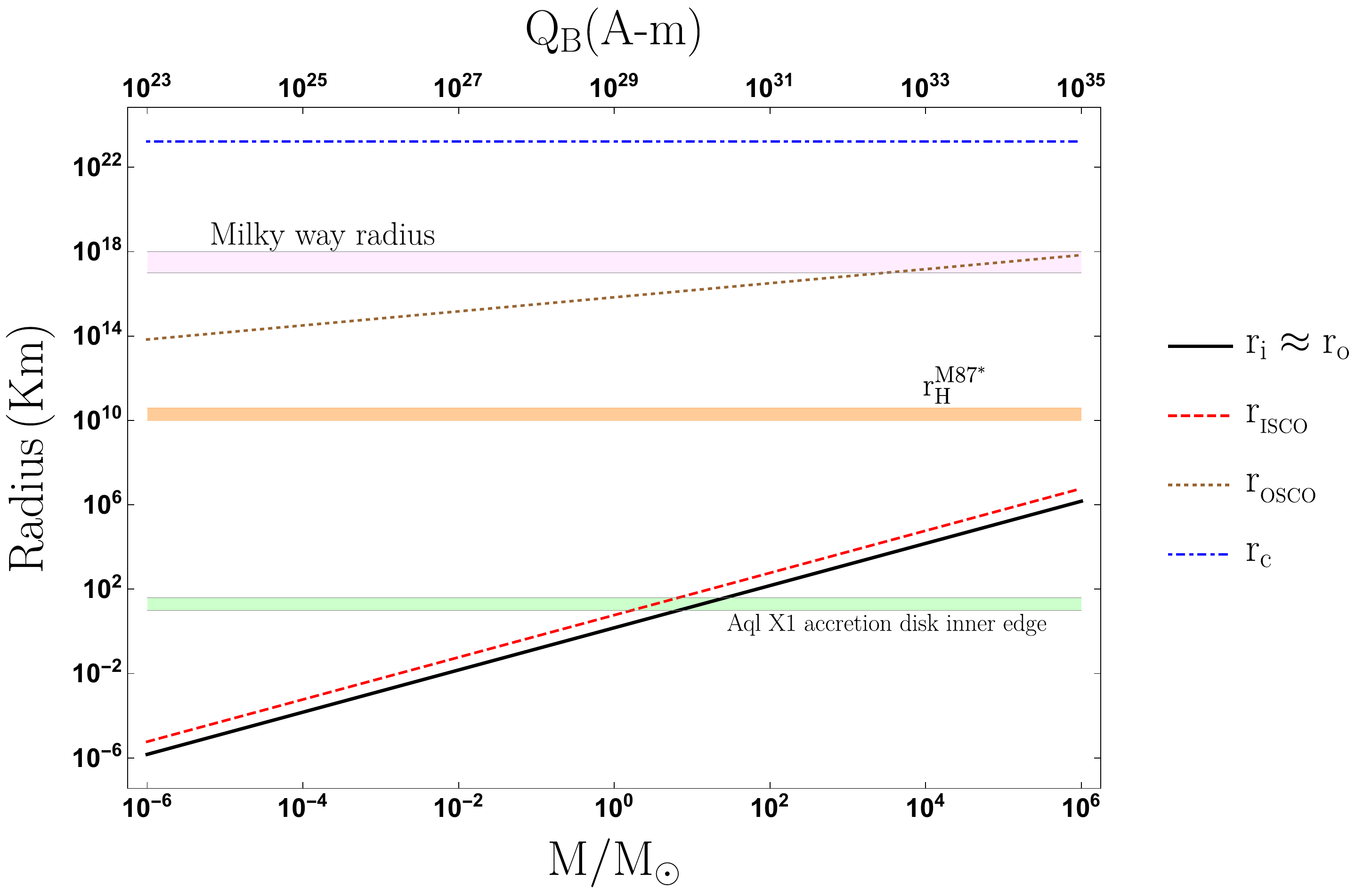}
\end{tabular}
\caption{Various length scales associated with an extremal MBH. The ISCO furnishes a characteristic length scale for various astrophysical phenomena, such as suggestively marking the edge of accretion discs. During binary inspiral of two compact objects, they also characterise typical separations before which the objects will begin to plunge into each other. In asymptotically de Sitter spacetime, an OSCO also appears which is displayed in the figure. For comparison, a few astrophysical length scales are overlaid---the radius of the Milky way galaxy, the indicative horizon radius for the M87* BH as measured by the Event Horizon Telescope\,\cite{Akiyama:2019cqa}, and the inner edge of the accretion disc measured for the Aquila X-1 neutron star system\,\cite{King_2016}.}
\label{fig:rH-vs-M}
\end{figure}
%%%%%%%%%%%%%%%%%%%%%%%%%%%%%%%%%%%%

As $r\rightarrow 2 |Q_B|$, $l^2$ starts to blow up, and hence illustrates the possible existence of an ISCO in the vicinity. With $Q_B^2 \Lambda \ll 1$, we suspected from Eq.\,(\ref{eq:rnisco}) that it must be close to $4 |Q_B|$. Iteratively solving, we already established the position of this ISCO in Eq.\,(\ref{eq:rndsisco}).

Focusing on the quartic in the numerator of Eq.\,(\ref{eq:rndslsq}), we deduce from Sturm's theorem (see for instance \cite{Barnard:1981}) that there are exactly two real roots; also, since coefficients of the quartic are real, the two imaginary roots come in complex conjugate pairs. From Descarte's rule of signs\,\cite{Barnard:1981}, we also further note that there are zero negative roots and therefore exactly two positive roots. 

Since again $Q_B^2 \Lambda \ll 1$, one of these roots must be very near $|Q_B|$. We are interested in the other positive root, in the context of finding an outer stable or quasi-stable circular orbit. A relatively sharp bound by Lagrange and Zassenhaus (see for instance\,\cite{Ostrowski:1960, knuth97}) for the positive roots of a polynomial $p(x)=a_n x^n+a_{n-1} x^{n-1}\ldots a_1 x+a_0$  is given by
\begin{equation}
x^*~\leq~2\,\mathrm{max} \left \{ \left| \frac{a_{n-1}}{a_n} \right|,   \left| \frac{a_{n-2}}{a_n} \right|^{1/2}\dots,   \left| \frac{a_{0}}{a_n} \right|^{1/n} \right\} \; .
\end{equation}
Thus, the larger of the positive roots for the numerator of Eq.\,(\ref{eq:rndslsq}), and hence the position of the OSCO, is bound by
\begin{equation}
r_{\text{\tiny{MBHdS}},m}^{\text{\tiny{OSCO}}} ~\leq~ 2\,\mathrm{max} \left \{ \left| \frac{3 Q_B}{\Lambda} \right|^{1/3}, \left| \frac{3 Q_B^2 }{\Lambda} \right|^{1/4} \right\} \; .
\end{equation}
It is found numerically that the bound is relatively good and gives an approximate analytic expression for the neighborhood of the OSCO. This supposition is also further strengthened by application of Sturm's theorem in ever tighter intervals. Due to the smallness of $\Lambda$, or more precisely due to $Q_B^2 \Lambda \ll 1$, the orbit corresponding to the largest positive root will be stable or at least quasi-stable with a very long lifetime.

In Fig.\,\ref{fig:rH-vs-M} we compare the various horizon scales and stable orbits of an MBH. The ISCO lies very close to the event horizon, and as we shall discuss in the next section, will have a tremendous magnetic field in its neighbourhood. With an increase in the MBH mass, the OSCO and ISCO separation gradually reduces. A few astrophysical length scales are also shown for comparisons.

%%%%%%%%%%%%%%%%%% Section %%%%%%%%%%%%%%%%%%%%%%%%%%%
\section{Quantum field theoretic and phenomenological aspects of MBH}
\label{sec:ewsr}
%%%%%%%%%%%%%%%%%% Section %%%%%%%%%%%%%%%%%%%%%%%%%%%

In this section we briefly review few relevant quantum field theoretic aspects of MBH and then some consequent constraints. We will focus on aspects of the near horizon magnetic fields, leading to symmetry restoration, Hawking radiation from extremal and near-extremal MBHs, and bounds on MBH dark matter from galactic magnetic field measurements. 

Let us consider the magnetic fields sourced by MBHs near the horizon and close to stable orbits in their vicinity. The magnetic field at the event horizon of an MBH  is given by
\begin{eqnarray}
B(r=r_H) &=&\frac{\mu_0}{4 \pi} \frac{1}{M} \frac{\frac{Q_B}{M}}{\left(1+ \sqrt{1- \left(\frac{Q_B}{M}\right)^2}\right)^2}\\
&=& 2.34 \times 10^{15} \, {\rm Tesla} \,  \frac{M_\odot}{M} \, \frac{\frac{Q_B}{M}}{\left(1+ \sqrt{1- \left(\frac{Q_B}{M}\right)^2}\right)^2}\nn \\
&\leq&  2.34 \times 10^{15} \, {\rm Tesla} \,  \frac{M_\odot}{M}\nn \, ,
\end{eqnarray}
where the equality holds for an extremal MBH, i.e., when $M =|Q_B|$. In the extremal case, the event horizon is located at $r_H=|Q_B|$. They fall inversely as the mass of the MBH. In Fig.~\ref{fig:B-vs-M}, we show variation of the magnetic field at the horizon and at the ISCO as a function of the MBH mass. Remarkably, even in mass ranges where no electroweak corona forms\,\cite{Maldacena:2020skw,Ambjorn:1988tm,Ambjorn:1989sz,Ambjorn:1989bd,Ambjorn:1992ca} , or equivalently electroweak symmetry is restored, the magnetic fields near the horizon and at the location of ISCO are immense. 

%%%%%%%%%%%%%%%%%%%%%%%%%%%%%%%%%%%%
\begin{figure}[!ht!]
\begin{center}
\begin{tabular}{cc}
\multicolumn{2}{c}{\includegraphics[scale=0.35]{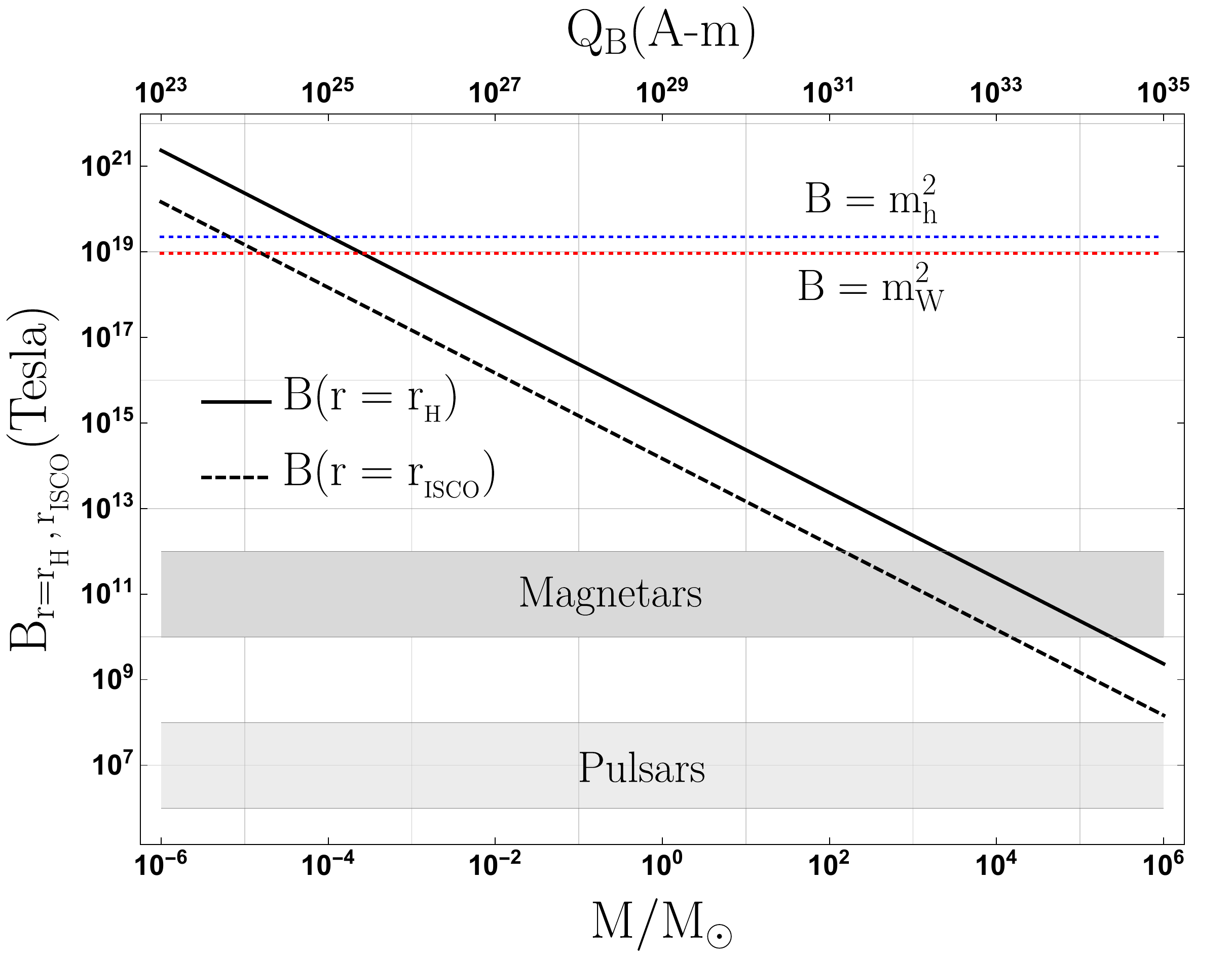}} 
\end{tabular}
\caption{ Magnetic field at the horizon and at the ISCO of an extremal MBH. The horizontal grey band corresponds to the typical magnetic field of a Magnetar\,\cite{Mereghetti:2008je}. It is observed that for a large range of MBH masses, the magnetic fields are very large near the horizon and at the location of the ISCO; even rivalling the currently know strongest cosmic magnets---Magnetars\,\cite{Mereghetti:2008je}. This opens the possibility that the MBH neighbourhoods may be very promising as probes of strong field quantum field theoretic phenomena\,\cite{Stohlker2003,Miransky:2015ava}. As we will also discuss, in Sec.\ref{sec:ememissions}, these large fields could also potentially lead to striking electromagnetic emissions.}
\label{fig:B-vs-M}
\end{center}
\end{figure}
%%%%%%%%%%%%%%%%%%%%%%%%%%%%%%%%%%%%

For a better perception of the magnitude of these MBH generated fields, across various masses, we may compare it to other large fields currently observed in the universe. Magnetars are a subset of neutron stars with extremely large magnetic fields---
\begin{equation}
B_{\text{\tiny{Magnetar}}} \gtrsim 10^{9}\,\mathrm{Tesla}\; .
\end{equation}
They are among the strongest cosmic magnets that are currently known\,\cite{Mereghetti:2008je}. We therefore note that the MBH fields near the event horizon, near the ISCOs, and even farther out, are generically much larger than typical Magnetar external fields. Hence, if MBHs exist in our universe, their neighbourhoods, where presumably accretion phenomena may be operational or trapped particles may exits, would be exquisitely suitable for testing strong field quantum field theoretic phenomena (see for instance\,\cite{Stohlker2003,Miransky:2015ava}). Apart from this, these immense fields, also make MBH interactions with neighbouring plasma or other compact objects, sources for striking electromagnetic phenomena. In Sec.\ref{sec:ememissions}, we will discuss a few aspects in this context.

Also, as discussed in \cite{Ambjorn:1992ca,Ambjorn:1989sz,Ambjorn:1989bd}, in the presence of a large magnetic field (larger than $m_h^2$), the electroweak symmetry may be restored and only the $\rm U(1)_{Y}$ component of the magnetic field survives; here Y denotes the Hypercharge. In this case, as emphasised in \cite{Maldacena:2020skw}, the near horizon region of the MBH will be in an electroweak symmetry unbroken phase, with intriguing properties. In order for the magnetic field to satisfy $B(r=r_H) \geq m_h^2 (\equiv 2.25 \times 10^{19} \, {\rm Tesla})$, a necessary condition is then to have $M \leq 1.04 \times 10^{-4} M_{\odot}$. For extremal MBHs, this would also correspond to $Q_B \leq 5.3 \times 10^{24} \rm \text{A-m}$; for a given $M$, the maximum magnetic field is obtained when $M=|Q_B|$. As one moves radially away from the horizon, the magnetic field keeps decreasing. The electroweak symmetry wll remain unbroken as long as $B(r=r_H)  \geq m_h^2$.  In the region where $B(r=r_H)  \leq m_W^2 (\equiv 0.93 \times 10^{19} \, {\rm Tesla})$, the electroweak symmetry is subsequently broken. 

In the intermediate region, dubbed the electroweak corona\,\cite{Maldacena:2020skw}, where $m_W^2 \leq B(r=r_H)  \leq m_h^2$, there is W-boson condensation and the Higgs vacuum expectation value is smaller than 246 GeV\, \cite{Ambjorn:1992ca,Ambjorn:1989sz,Ambjorn:1989bd}. The characteristic radius of the electroweak corona is
\begin{equation}
r_{\text{\tiny{corona}}}\sim \sqrt{\frac{Q_B}{2q_B \,m_W\,m_H}} \; .
\end{equation}
In the relevant low-mass range of Figs. \ref{fig:rH-vs-M} and \ref{fig:B-vs-M} and, where an electroweak corona may form, one has $r_{\text{\tiny{corona}}} \sim \mathcal{O}(1)\,\mathrm{mm}$ and decreasing as $\sqrt{M}$ for the smaller mass ranges. For the higher mass-ranges in this region of parameter space, one therefore has a macroscopic region extending outside the horizon. This region where electroweak symmetry is restored may affect processes and emissions close to the horizon, during accretion, just before the in-falling matter enters the event horizon. The Event Horizon Telescope may offer an opportunity to probe such potential near-horizon phenomena in the future\,\cite{Akiyama:2019cqa}.

 The electroweak corona and electroweak hair surrounding such low-mass MBHs may have interesting features, such as being non-spherically symmetric\,\cite{Lee:1994sk,Maldacena:2020skw}. These different regions of MBH mass are pictorially shown in Fig.~\ref{MBH-mass-regions}. A comprehensive exploration of the phenomenology of MBHs, with electroweak coronas, in this low-mass region is presented in\,\cite{Bai:2020spd}.

Let us now briefly discuss a few relevant aspects of MBH thermodynamics. The temperature of a MBH is given by
\begin{eqnarray}
T(r=r_H)  &=& \frac{1}{8 \pi M} \frac{4 \sqrt{1- \left(\frac{Q_B}{M}\right)^2} }{\left(1+
\sqrt{1- \left(\frac{Q_B}{M}\right)^2}\right)^2}\\
&=& 6.15 \times 10^{-8} \, {\rm K} \, \frac{M_\odot}{M}   \, \frac{4 \sqrt{1- \left(\frac{Q_B}{M}\right)^2} }{\left(1+
\sqrt{1- \left(\frac{Q_B}{M}\right)^2}\right)^2} \nn \\
&=& 5.3 \times 10^{-18} \, {\rm MeV} \, \frac{M_\odot}{M}   \, \frac{4 \sqrt{1- \left(\frac{Q_B}{M}\right)^2} }{\left(1+
\sqrt{1- \left(\frac{Q_B}{M}\right)^2}\right)^2}   \nn \\
&\leq&  \frac{1}{8 \pi M} \, . \nn
\end{eqnarray}

The temperature is zero for an extremal MBH. 
Due to the existence of the near-horizon electroweak symmetric region, and thus massless fermions,
the Hawking radiation from a MBH can be modified. 
It was shown in \cite{Maldacena:2018gjk} that a 3+1 dimensional massless chiral fermion with charge $q$ (in the current
context, $q$ would be the Hypercharge) gives rise to $q Q_B/q_B$ (where again as before we define $q_B = 2 \pi \epsilon_0 \hbar c^2/e = 3.44 \times 10^{-9}$ A-m) massless 1+1 dimensional chiral fermions. The vanishing mass may be understood by noting that the lowest Landau levels for fermions in the presence of magnetic field have zero energy. This means that the radiated power will be
proportional to $(q Q_B/q_B) T^2$, which is the same temperature dependence as for a blackbody radiation in 1+1 dimension. This is bigger by a factor $q Q_B$ than the power radiated from a Schwarzschild black hole of the same horizon size. Thus, there is increased radiation as compared to a neutral black hole. 

For example, consider that the MBH temperature satisfies $T>m_e$, which corresponds to $M \lesssim 1.03 \times 10^{-17} M_{\odot}$.
The Hawking radiation will be enhanced by the above degeneracy factor. Now, even for $T <m_e$, emission of electrons is modified in the near-horizon
electroweak symmetric region, since electrons are massless there. However, these electrons cannot escape to infinity since they have non-zero
mass in the region with $B < m_W^2$\cite{Maldacena:2020skw}. As also discussed in\,\cite{Maldacena:2020skw}, due to existence of the 1+1 dimensional modes, a non-extremal Black Hole with $T>m_e$ (and $B(r=r_H) \geq m_h^2 $) would rather quickly radiate away energy to become an extremal black hole. Thus, apart from some of the astrophysical signatures we discuss in this work, low-mass MBH where an electroweak corona forms, may also have additional phenomenological and astrophysical features\,\cite{Bai:2020spd}.

%%%%%%%%%%%%%%%%%%%%%%%%%%%%%%%%%%%%
\begin{figure}[!ht!]
\begin{center}
\begin{tabular}{c}
\hspace{-5mm}\includegraphics[scale=0.275]{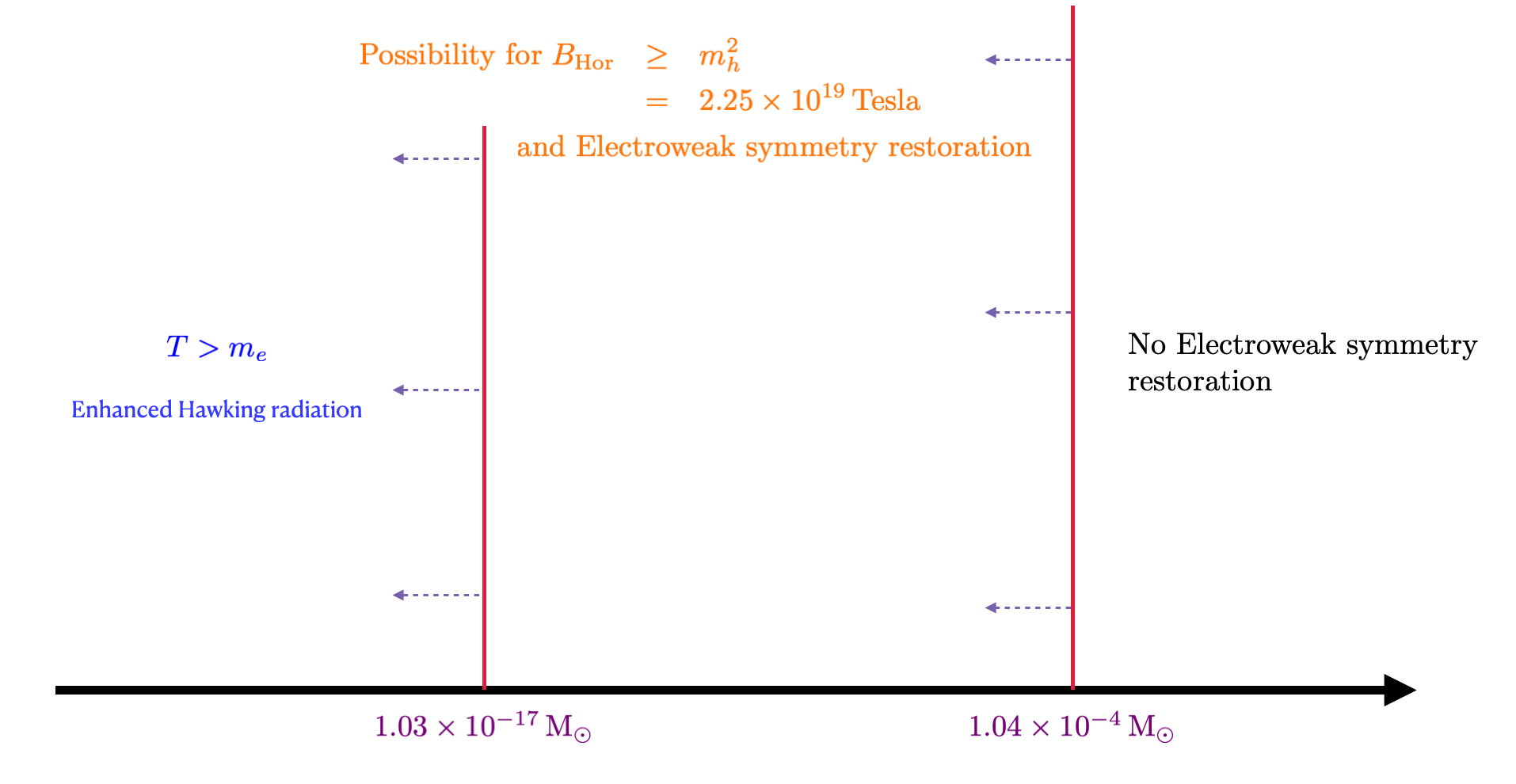}
\end{tabular}
\caption{Different regions of MBH mass with qualitatively different features as discussed in the text. For $M \gtrsim 10^{-4} M_\odot $, the magnetic field at the Horizon is not large enough to
render electroweak symmetry restoration. For $M \le 10^{-4} M_\odot $, the electroweak symmetry can in principle be restored near the Horizon depending on the magnetic charge $Q_B$.
Furthermore, for $M \le 10^{-17} M_\odot $ the Hawking radiation rate is enhanced due to the emergence of massless 1+1 dimensional fermionic modes with large degeneracy
which can escape to a large distance., see text for more details.}
\label{MBH-mass-regions}
\end{center}
\end{figure}
%%%%%%%%%%%%%%%%%%%%%%%%%%%%%%%%%%%%

Finally, let us discuss a few pertinent aspects of a scenario where MBHs may be speculated to be a component of the dark matter (DM) in our universe. Primordial neutral black holes are a popular candidate for DM, see \cite{Carr:2020xqk} for a recent review.  Thus, it is interesting to also consider the possibility of primordial MBHs as dark matter candidate. However, astrophysical bound
on such a possibility can be quite stringent. One such bound arises from that fact that MBHs will drain energy from the galactic
magnetic field when they pass through it, and thus, requiring the survival of such magnetic fields today can give strong constraints
on the abundance of MBHs. This general line of reasoning leads to the so called Parker bound \cite{Parker:1970xv,Turner:1982ag,Adams:1993fj}. 

Let us discuss the Parker bound, in our context, in some detail. Specifically, we will derive the bound on the local density of MBHs following the analyses in\,\cite{Parker:1970xv,Turner:1982ag,Adams:1993fj}. For the astrophysical parameters we adopt, our results agree with the treatment in\,\cite{Turner:1982ag} but are slightly different from\,\cite{Bai:2020spd}, where the focus was on MBHs with an electroweak corona and masses below $10^{-4} M_{\odot}$. The results nevertheless are consistent within astrophysical uncertainties and other modelling assumptions. 

Consider the characteristic speed 
\begin{equation}
v_{mag}=\sqrt{\frac{2BQl_c}{M}}\;.
\end{equation}
This is the speed a monopole, starting from rest, gains in a constant magnetic field after traversing unit coherence length $l_c$.  $l_c$ is the characteristic length of the galactic domain over which the magnetic field may be assumed to be roughly constant. In order for the MBHs to constitute the DM of our galaxy, their average velocity near the solar system should be $v\sim 10^{-3} c $ (assuming $\rho_{DM}^{\rm local} \approx 0.4 \rm \, GeV/cm^3$) and $v\gg v_{mag}$. This is because, if we are assuming that the MBHs constitute the DM of the galaxy, their velocity must be smaller than the galactic virial velocity to remain bound to the galaxy. Also, if $v\sim 10^{-3}c \ll v_{mag}$, then the monopole will quickly accelerate to $v_{mag}$ and eventually leave the galaxy. This clearly is not the case we are interested in. 

Assume that a MBH with charge $Q_B$ and mass $M$ enters a region which has a constant magnetic field $\vec{B}$ over a coherence length $l_c$. Due to the magnetic Lorentz force, the change in kinetic energy of the MBH would be equal to $\frac{Q_B^2 B^2 l_c^2}{2mv^2}$. However, the change in the magnitude of the velocity due to the
magnetic force ($\Delta v$) must be such that it satisfies $\Delta v \ll v$, since we are assuming the $v \gg v_{mag}$ regime. This then requires the MBH to satisfy
\begin{equation}
\frac{Q_B}{M} \ll   15  \, \rm \frac{\text{A-m}}{Kg} \; .
\end{equation}
This condition is always satisfied by a near-extremal or extremal MBH with $Q_B=M$; or equivalently, in SI units, satisfying $Q_B/M \simeq 2.6 \times 10^{-2} \rm \text{A-m}/Kg$.

The equation of motion for a MBH moving through constant magnetic field is given by
\begin{eqnarray}
  \frac{d\vec{v}}{dt} &=& \frac{Q_B\vec{B}}{M} \; ,\nn \\
 \implies\frac{dE_k}{dt} &=& Q_B\vec{B}.\vec{v} \, .
 \label{eq-A3}
\end{eqnarray}
Here, $E_k$ is the kinetic energy of the black hole, Assuming an isotropic velocity distribution, one has $\big{<}\vec{B}.\vec{v}\big{>}=0$, implying $\big{<}\frac{dE_k}{dt}\big{>}=0$.

Now, differentiating Eq.~\eqref{eq-A3} with respect to time, one gets
\begin{equation}
    \frac{d^{2}E_k}{dt^{2}}=\frac{Q_B^{2}B^{2}}{M}>0 \; .
\end{equation}
One may therefore write, 
\begin{eqnarray}
    \big{<}E_k(t+\Delta t)\big{>} &\simeq& \big{<}E_k(t)\big{>}+\frac{1}{2}(\Delta t)^{2}\big{<}\frac{d^{2}E_k}{dt^{2}}\big{>}\; ,\nn\\
   \implies \Delta E_k&\simeq&(\Delta t)^{2}\frac{Q_B^{2}B^{2}}{2M}=\frac{Q_B^{2}B^{2}l_c^{2}}{2Mv^{2}} \; .
\end{eqnarray}
Here $\Delta t=l_c/v$, and we are assuming that the MBH is deflected only slightly. Hence, by looking at the second order contribution, it is clear that the kinetic energy increases on the average, although the  first order contribution is vanishing.

Assuming that the MBHs account for a faction $f_{\rm DM}$ of the observed local DM density, their flux may be approximated as
\begin{eqnarray}
            F_{\rm MBH} &=&  \frac{2.13 \times 10^{-16}}{m^2  \, sec.}  \, f_{\rm DM} \bigg{(}\frac{\rm Kg}{M} \bigg{)}
            \bigg{(}\frac{v}{10^{-3}c}\bigg{)}  \nn \\
            &\times &\bigg{(}\frac{\rho_{\rm DM}}{ 0.4 \, \rm GeV/cm^3}\bigg{)} \; .
            \label{eq-A7}
 \end{eqnarray}
  
The Parker bound\,\cite{Parker:1970xv,Turner:1982ag,Adams:1993fj} is obtained by requiring that the average energy gained by the MBHs, during the regeneration time $t_{reg}$ of the galactic magnetic field, be smaller than the energy stored in the magnetic field. From this line of argument, one gets
 \begin{equation}
  \Delta E_k *F_{\rm MBH}*(4\pi l_c^{2})*t_{reg} \leq \bigg{(}\frac{1}{2\mu_0}B^{2}\bigg{)}\bigg{(} \frac{4}{3}\pi l_c^{3} \bigg{)} \; ,
 \end{equation}
 which implies that the flux is bound by 
 \begin{equation}
F_{\rm MBH}  \leq \frac{M v^{2}}{3 \mu_0Q_B^{2}l_c t_{reg}} \; .
\label{eq:fluxbound}
\end{equation}

Using Eq.\,(\ref{eq:fluxbound}) now in Eq.\,\eqref{eq-A7}, we get a limit on the MBH DM fraction
\begin{eqnarray}
f_{\rm DM} &\leq& 1.5 \times 10^{-6} \left( \frac{M}{\rm Kg} \right)^2  \left( \frac{\text{A-m}}{Q_B} \right)^2 \left( \frac{v}{10^{-3} c} \right) \nn \\
&\times&
\left( \frac{ 0.4 \, \rm GeV/cm^3}{\rho_{\rm DM}} \right) \left( \frac{10 \text{kpc}}{l_c} \right) \left( \frac{ 10 \text{Gyr}}{t_{reg}} \right) \; .
\end{eqnarray}
Assuming typical astrophysical parameter values, the fraction of DM, $f_{\rm DM}$, that can be made of the MBHs, is therefore constrained to be around 
\begin{align}
f_{DM} \lesssim 1.5 \times 10^{-6} \left(\dfrac{M/\text{Kg}}{Q_B/\text{A-m}}\right)^2 \; .
\end{align}

For an extremal MBH, which has 
\begin{equation}
\left(\dfrac{M/\text{Kg}}{Q_B/\text{A-m}}\right)  = 38.7
\end{equation}
one gets 
\begin{equation}
f_{DM} \lesssim 1.7 \times 10^{-3}  \; .
\end{equation}
A near-extremal MBH, as a DM candidate would be subject to a weaker bound. For instance, if there is say a $10\%$ deviation from extremality,
\begin{equation}
\left(\dfrac{M/\text{Kg}}{Q_B/\text{A-m}}\right)  = 1.1 \times 38.7 \; ,
\end{equation}
the bound would be approximately $20\%$ weaker
\begin{equation}
f_{DM} \lesssim 2.1 \times 10^{-3} \; .
\end{equation}
These indicative bounds show that, unlike the uncharged primordial BHs (in certain mass ranges), it is very unlikely that the extremal or near-extremal MBHs constitute a very significant fraction of DM in our universe. 

Note that the above bounds on the DM fraction are derived assuming parameter values corresponding to Andromeda galaxy (M31), for which $l_c = 10 \, \rm kpc$, $t_{reg} = 10 \, \rm Gyr$, $\rho_{\rm DM} = 0.4 \, \rm GeV/cm^3$ and $v = 10^{-3}c$\,\cite{Fletcher:2003ec,Arshakian:2008cx,Klypin:2001xu,stellarMass}. The possibility to strengthen the Parker bound by using M31 magnetic field measurements was first pointed out in\,\cite{Bai:2020spd}. Improvements in galactic magnetic field measurements opens up the possibility that this bound may be further strengthened in future\,\cite{2011AIPC.1381..117B}.
 
 Before we conclude this section, we would like to very briefly comment on the possibility of nucleon decay catalysis
by the MBHs, similar to the Rubakov-Callan effect for magnetic monopoles\,\cite{Rubakov:1981rg, Rubakov:1982fp, Callan:1982ah, Callan:1982au, Callan:1982ac,Sen:1984qe}. We like to note here that magnetic monopoles need not always catalyse proton decay. It depends on whether the physics at the core of the monopole violates baryon number or not. It should however be noted that already the Standard Model weak interaction do violate baryon number, by the triangle anomaly, and can in principle lead to unsuppressed proton decay catalysis\,\cite{Sen:1984qw,Sen:1984ku}. In the original proposal by Rubakov and Callan, it was the baryon number violating gauge field configuration inside the monopole core which was responsible for the baryon number violation. In fact, soon after Rubakov and Callan's work, examples were constructed where monopoles indeed did not catalyse nucleon decay\,\cite{Dawson:1982sc,Weinberg:1983bf}. Moreover, a Rubakov-Callan like analysis in the background of a MBH may have more subtleties\,\cite{Maldacena:2020skw} due to the fact that the magnetically charged object now is a black hole, with an event horizon. Thus, to our understanding, no robust and model independent bound can be derived at present on the MBH abundance based on
Rubakov-Callan like effect.

 %%%%%%%%%%%%%%%%%% Section %%%%%%%%%%%%%%%%%%%%%%%%%%%
\section{Electromagnetic emissions from MBH binary inspirals}{
\label{sec:ememissions}

%%%%%%%%%%%%%%%%%%%%%
If an MBH is involved in a binary inspiral, due to the magnetic charge, one expects electromagnetic emissions from the system. We would like to estimate the power radiated as well as understand some characteristics of this electromagnetic emission, for typical extremal MBH parameters. We will consider two cases-- where only one of the black holes in the binary is an extremal MBH, and the case where both black holes in the binary are extremal MBHs. 

Let us consider the case of a neutral black hole and an extremal MBH undergoing binary inspiral (MBH-BH case). At sufficiently large separations (relative to event horizon radii), during an epoch long before the actual merger of the black holes, we may neglect corrections to Maxwell's equations due to spacetime curvature, as a first approximation. If $\dot{\omega}/\omega^2 \ll 1$, where $\omega$ is the orbital angular velocity, the orbits may be further approximated as an adiabatic sequence of quasi circular ones. This is well satisfied in the earlier phase of the binary inspiral. The electromagnetic and gravitational radiation reactions also tend to make the orbits circular\,\cite{Maggiore:1900zz}, thus making the assumption of quasi-circular orbits well motivated. All these then imply that the instantaneous acceleration and velocity are also well approximated to be perpendicular to each other. Finally, to focus just on the essentials, we will also take $Q_{B,1}=Q_{B,2}\equiv Q_B$ for the extremal black holes. These well motivated assumptions, as we shall see, will enable us to gain a clearer analytic understanding of the electromagnetic emissions from the MBH-BH system.

With the above assumptions, we may use the Li\'enard generalization of the Larmor formula to estimate the differential power radiated per solid angle for MBH-BH inspiral (see appendix \ref{app-B}). To estimate the radiation pattern and the net power emitted it suffices to consider the situation at a particular instant. The instantaneous velocity and acceleration of the MBH are perpendicular to each other and the radiated power per solid angle for the MBH-BH may be computed as (appendix \ref{app-B})
\begin{eqnarray}
\frac{dP^{\text{\tiny{MBH-BH}}} }{d\Omega}&&= \frac{\mu_0 Q_B^2 a^2 \omega^4}{64\pi^2c^3}  \cdot \\
&& \frac{\left[\left( 1-\beta \sin\theta' \sin\phi' \right)^2- \left(1-\beta^2\right)\sin^2\theta' \cos^2\phi' \right]}{\left( 1-\beta  \sin\theta' \sin\phi' \right)^5} \; .\nn
\end{eqnarray}
$\beta=(\omega a)/2c$ is the boost and $a$ is the separation between the MBH and BH at that instant. The prime denotes that we are adopting a  coordinate frame of reference attached to the MBH and that the angles are with respect to it.

When $\beta \ll 1$, during the early inspiral phase, most of the radiation is emitted normal to the orbital plane.
\begin{equation}
\frac{dP^{\text{\tiny{MBH-BH}}} }{d\Omega}\Big\vert_{\beta \ll 1} \propto 1-\sin^2\theta' \cos^2\phi' \; .
\end{equation}

%%%%%%%%%%%%%%%%%%%%%%%%%%%%%%%%%%%%
\begin{figure}
\begin{center}
\begin{tabular}{c}
\includegraphics[scale=0.3]{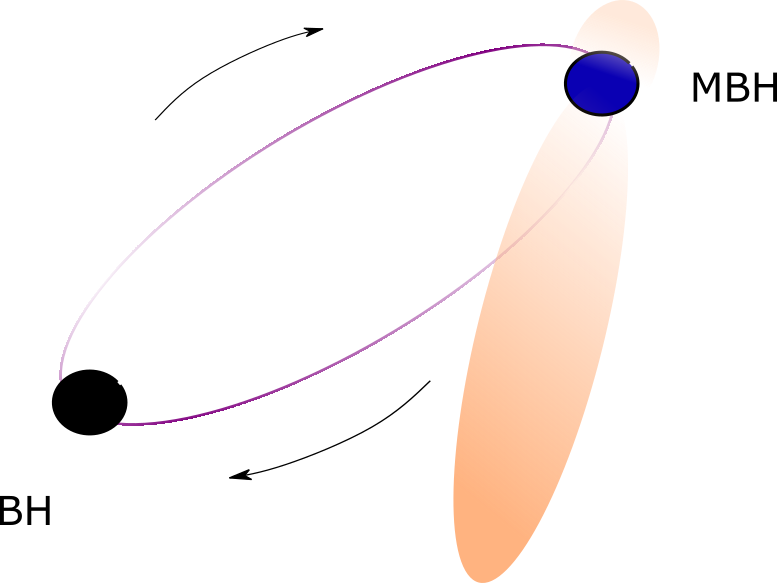}
\end{tabular}
\caption{While for $\beta \ll 1$, most of the electromagnetic Poynting flux is directed perpendicular to the orbital plane, as  $\beta \rightarrow 1$ the radiation profile qualitatively changes. Most of the emitted electromagnetic power for $\beta\rightarrow 1$ is in the orbital plane, and points in the direction of the MBH's instantaneous velocity. If an observer is in the line of sight of this beam, they would see a quasi-periodic emission that gradually increases in frequency. Also, as we will quantify, if MBHs exist and undergo capture by another compact object to form an MBH-BH binary pair or something similar, the total electromagnetic power output is generically very large---rivalling even some of the most energetic Gamma Ray Bursts known to date (see discussions in the text and Fig.\,\ref{fig:em-power}).}
\label{fig:MBH-BH-EMprofile}
\end{center}
\end{figure}
%%%%%%%%%%%%%%%%%%%%%%%%%%%%%%%%%%%%

As the orbital speeds increase and $\beta\rightarrow 1$, later in the inspiral evolution, the radiation profile changes qualitatively. One notes that the radiation is now dominantly emitted in the direction of the MBH's instantaneous velocity.
\begin{equation}
\frac{dP^{\text{\tiny{MBH-BH}}} }{d\Omega}\Big\vert_{\beta \rightarrow1} \sim \frac{1}{(1-\sin\theta' \sin\phi')^3} \; .
\end{equation}
The radiation profile is dominant in the forward direction to the instantaneous velocity and symmetric about  it. The electromagnetic emission sweeps forward like a headlight, as the MBH orbits (see Fig.\,\ref{fig:MBH-BH-EMprofile}). 

This is of course just analogous to synchrotron radiation from an electrically charged particle moving in a circle. Thus, if earth is in the path of this intense beam at some point in the orbit, we may be able to detect this large and quasi-periodic Poynting flux as it repeatedly sweeps by. Fast Radio Bursts\,\cite{Lorimer:2007qn,Thornton:2013iua} are mysterious, radio transients that last for $\mathcal{O}(1)\,\mathrm{ms}$, whose origins are not currently understood. It is intriguing to speculate if, at least in some cases, Fast Radio Bursts may be related to this sweeping of the electromagnetic emission, as an MBH undergoes orbital motion. Even when $\beta \ll 1$, the binary system would be driving a very energetic Poynting flux with ever increasing power. For electrically charged black holes some speculation in this general direction exists in the literature\,\cite{Zhang:2016rli}. There has also been interesting ideas proposed related to the collapse of magnetospheres in Kerr-Newman black holes\,\cite{Liu:2016olx}. The beaming effect in binaries involving MBHs, as $\beta \rightarrow 1$, nevertheless has not been speculated on before. Compared to electrically charged compact objects, for MBHs, the considerations may also be markedly different---due to their better astrophysical viability, potentially larger charges, as well as corresponding changes in the electromagnetic emissions and orbital evolutions.

%%%%%%%%%%%%%%%%%%%%%%%%%%%%%%%%%%%%
\begin{figure}[!ht!]
\begin{center}
\begin{tabular}{c}
\includegraphics[scale=0.3]{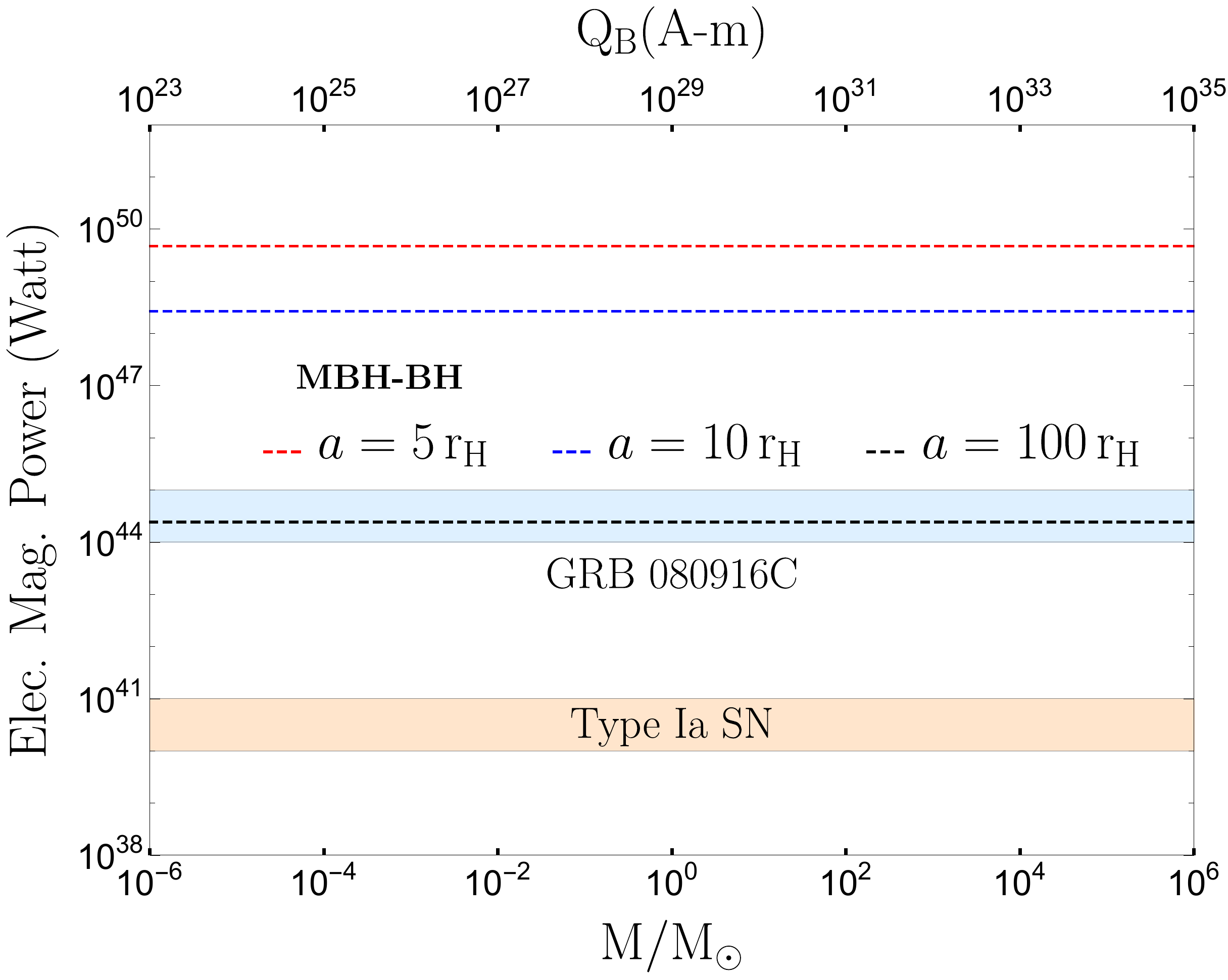} \\
 \includegraphics[scale=0.3]{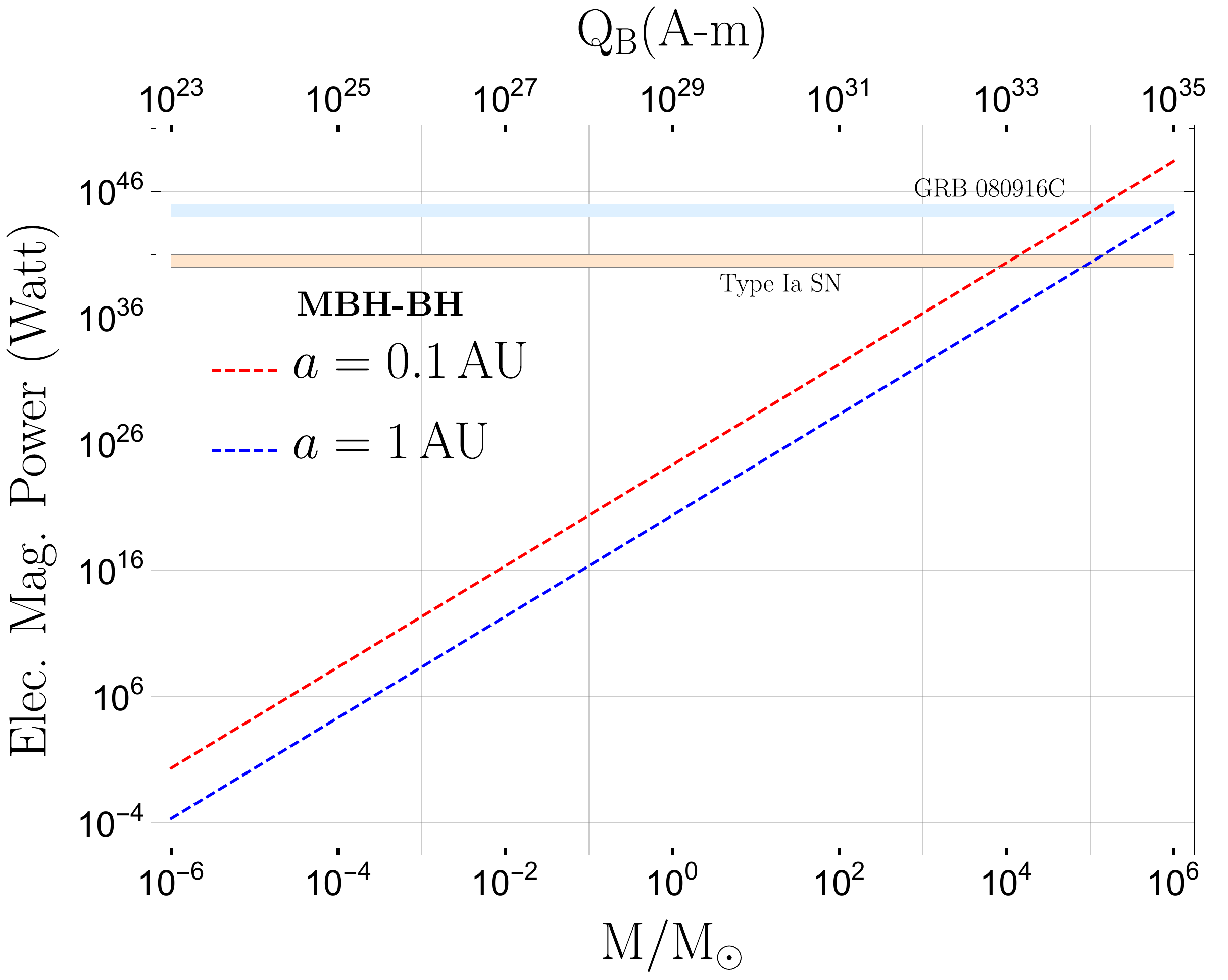}
\end{tabular}
\caption{The figure displays the total electromagnetic power emitted, during binary inspiral of a blackhole system where one of the objects is an extremal  MBH. For the MBH-MBH case, the total power emitted is just a factor of four larger. The emitted power is displayed as a function of the common black hole masses and the different curves are for different black hole separations. When the separation is quantified as a fixed multiple of the horizon length, the dependence on mass drops-out, due to the assumption of extremality. For comparison, we display typical power emitted by Type Ia supernovae, and note that the inspiral electromagnetic power expected from MBHs may be many orders of magnitude larger. Intriguingly, the emitted power may even rival some of the most powerful Gamma Ray Bursts observed in our universe--- for instance, GRB 080916C\,\cite{Abdo:2009zza} as overlaid on the figure.}
\label{fig:em-power}
\end{center}
\end{figure}
%%%%%%%%%%%%%%%%%%%%%%%%%%%%%%%%%%%%

The total power radiated may be computed from above to be
\begin{equation}
P^{\text{\tiny{MBH-BH}}} = \frac{\mu_0 Q_B^2 a^2 \omega^4}{24 \pi c^3}\; .
\end{equation}
To understand the above emitted power as a function of the MBH-BH separation or orbital angular velocity, one may use the Keplerian relation to express the orbital angular velocity also as a function of $a$, or vice versa. The total electromagnetic power emitted varies as $a^{-4}$ or equivalently as $\omega^{8/3}$. Now, the Lorentz factor defined as $\gamma=1/\sqrt{1- \frac{\omega^2 a^2}{4 c^2}}$ gradually increases as the binary inspiral evolves and closer to the merger phase one expects $\beta \rightarrow 1$. The emitted electromagnetic power in that regime is expected to be further enhanced by a factor $\sim \gamma^4$. This can lead to even larger electromagnetic emissions as the black holes inspiral closer to each other. The other effect is that the frequency of the emitted radiation will increase as the black hole separations decreases, evolving roughly as $\omega_{\text{\tiny{EM}}} \sim a^{-3/2}$, starting from the far radio band and progressing to gamma ray frequencies. In the late phase of the binary inspiral, the curvature effects will start becoming important and will also have to be gradually accounted for while treating electromagnetic emissions. Some aspects of these curvature corrections, for dyonic black hole inspirals, have been discussed in the literature\,\cite{Liu:2020vsy}. 

In Fig.\,\ref{fig:em-power} we compare the total electromagnetic power emitted assuming various separations. When the separation is quantified as a fixed multiple of the horizon radius, one sees that the emitted power is independent of $Q_B$ (or $M$) due to the assumption of extremality. This is because, when the Keplerian relation is used along with the expression for $r_H$, the dependence on $Q_B$ drops out. The power emitted by a few astrophysical sources are also displayed in the figure. Remarkably, one notes that the total power emitted in binary inspirals, involving an MBH, may be much larger than those from Type Ia supernovae and very energetic Gamma Ray Bursts.
 
 Let us now consider the second case of interest---the inspiral of two oppositely charged extremal MBHs (the MBH-MBH case). For simplicity and to focus on the salient aspects, we again take the magnitude of the charges to be the same for both the magnetic black holes. We also follow all the simplifying assumptions that we had adopted for the MBH-BH case. Now, as the two MBHS revolve around each other, they will again generate magnetic dipole radiation. The emitted Poynting flux is now expected to be larger than the MBH-BH case, owing to the fact that both objects are now magnetically charged. 
 
 Now, under the assumptions we are working with, the retarded potentials and electromagnetic fields due to the two revolving MBHs may be  approximated as that due to the superposition of two oscillating magnetic dipoles. The two dipoles just differ by a $\pi/2$ phase difference, under the assumption of quasi-circular orbits. The electromagnetic fields and the corresponding Poynting vector may be computed readily (see appendix\,\ref{app-B}), giving the instantaneous power radiated per unit area as
\begin{equation}
\vec{S}^{\text{\tiny{MBH-MBH}}} = \frac{\mu_0 Q_B^2 a^2 \omega^4}{16 \pi^2 c^3 r^2} \left[ 1-\left\{ \sin\theta \cos\left(\omega t_r -\phi \right)\right\}^2\right]~\hat{r} \; .
\end{equation}
 Here, $r$ is the distance from the binary system to the observer and $t_r=t-r/c$ is the retarded time. When averaged over a complete cycle, this gives the mean directional intensity of the emitted electromagnetic radiation in the MBH-MBH case,
\begin{equation}
I^{\text{\tiny{MBH-MBH}}}_{\text{\tiny{EW}}} ~\hat{r}=\langle \vec{S}^{\text{\tiny{MBH-MBH}}} \rangle =  \frac{\mu_0 Q_B^2 a^2 \omega^4}{16 \pi^2 c^3 r^2} \left[ 1-\frac{1}{2} \sin^2\theta\right]~\hat{r} \; .
\end{equation}
The maximum intensity is again directed normal to the orbital plane, for $\beta \ll 1$. It falls off as $1-\frac{1}{2} \sin^2\theta$ with increasing polar angle, with moderate emissions in the orbital plane. The generated radiation is found to be in general elliptically polarized, but normal to the orbital plane, along the direction of maximum intensity ($\hat{z}$), it will be circularly polarized.

Within our approximations, the total power radiated may now be estimated by integrating over a sphere surrounding the binary system. This gives
\begin{equation}
P_{\text{\tiny{EW}}}^{\text{\tiny{MBH-MBH}}}= \frac{\mu_0 Q_B^2 a^2 \omega^4}{6\pi c^3} \; .
\label{eq:empow}
\end{equation}
This is twice the power radiated by a single oscillating magnetic dipole. This is because, though quadratic in the fields, the cross terms with fields from each of the superposing dipoles average to zero, being out of phase by $\pi/2$. This is also four times greater than the total power emitted in the MBH-BH case. 

Again, as they spiral inward and the orbital speeds become relativistic, the total power radiated will be enhanced by a factor $\gamma^4$. Thus in the later epochs of the inspiral, the electromagnetic power radiated can be further enhanced. Also, for $\beta \rightarrow 1$, as in the MBH-BH case, most of the emitted power is expected to be in the orbital plane. The emitted electromagnetic frequency also increases from the far radio band to gamma ray bands, as the inspiral evolves, going again as $a^{-3/2}$. One point to note though is that the Keplerian relations are slightly modified in the MBH-MBH case---owing to the fact that for extremal MBH pairs, the electromagnetic attraction is exactly equal to the gravitational attraction. Hence, the actual numerical factors and coefficients relating $\omega_{\text{\tiny{EM}}}$ to orbital separation will be slightly different from the MBH-BH case.

For simplicity and to focus on the most salient aspects, we have assumed equal masses and equal magnitudes for the magnetic charges everywhere, but the expressions may be extended to cases where $M_1\neq M_2$ and $Q_{B,1}\neq Q_{B,2}$, as well as for motions with non-zero eccentricities (see for instance\,\cite{Landau:1982dva}). We do not expect the main points to drastically change from the present analysis nevertheless.

 These electromagnetic emissions and the associated radiation reactions will contribute to the gradual evolution of the orbit. Thus, to analyse the orbital and gravitational wave frequency evolution, unlike conventional binary inspirals with just neutral black holes, we will need to include these in addition to the power radiated by gravitational waves. We will investigate these aspects in the next section. There, we will analyse the modification to the orbital frequency evolution, and consequently the evolution of emitted gravitational frequencies, along with changes to the inter black hole separation.

%%%%%%%%%%%%%%%%%% Section %%%%%%%%%%%%%%%%%%%%%%%%%%%
\section{Gravitational waves from binary inspirals of MBH}{
\label{sec:gwemissions}
%%%%%%%%%%%%%%%%%%%%%

The dynamics of the MBH-BH or MBH-MBH binary systems will be modified from conventional expectations, due to the large additional electromagnetic Poynting flux. Apart from a possible electromagnetic counterpart, this must also manifest in the evolution of the orbits and the gravitational waveforms. We would like to investigate this in some detail, and we adopt the same assumptions as those made in the previous section, to obtain simple analytic results.

%%%%%%%%%%%%%%%%%%%%%%%%%%%%%%%%%%%%
\begin{figure}[!ht!]
\begin{center}
\begin{tabular}{cc}
\includegraphics[scale=0.18]{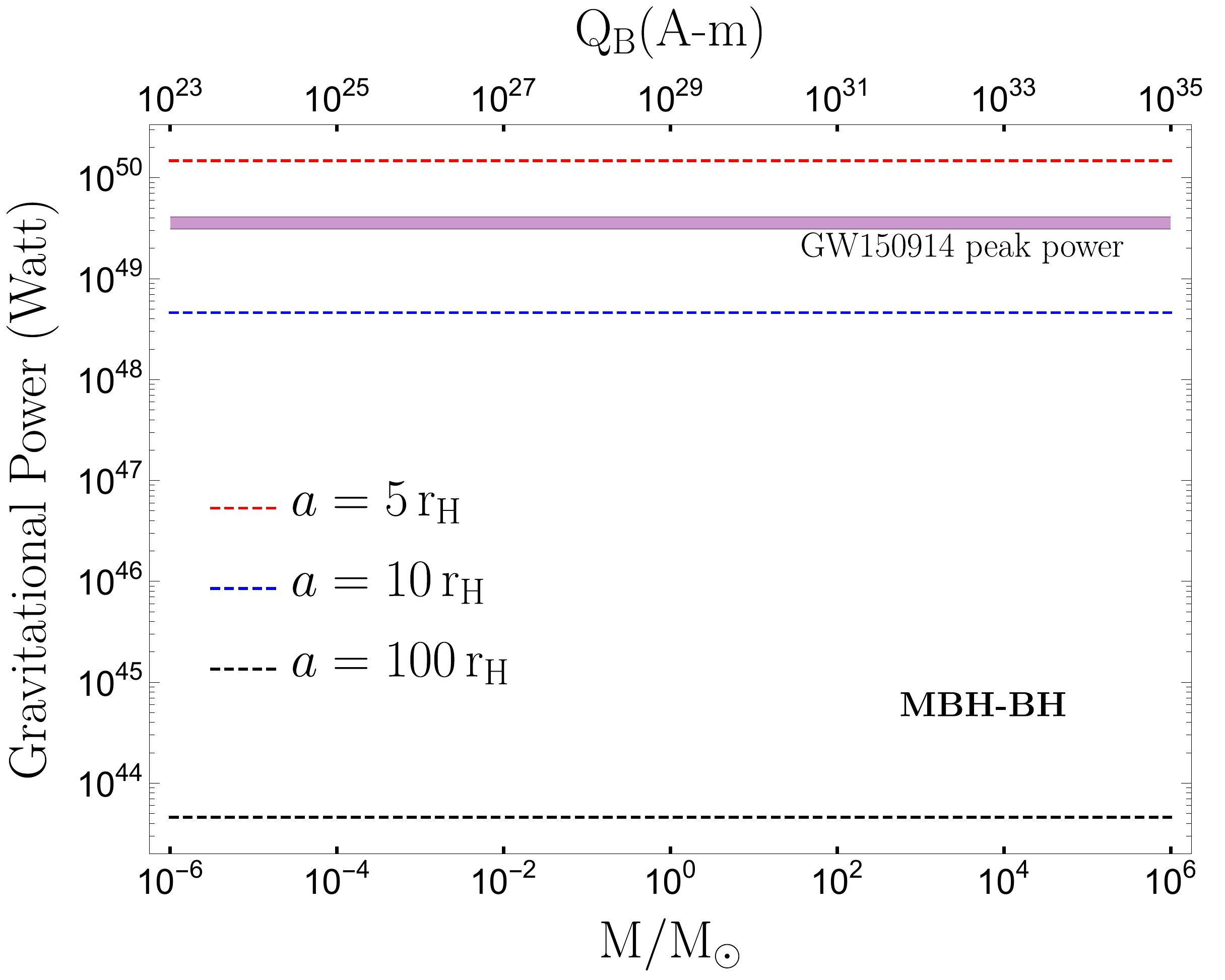} & \includegraphics[scale=0.18]{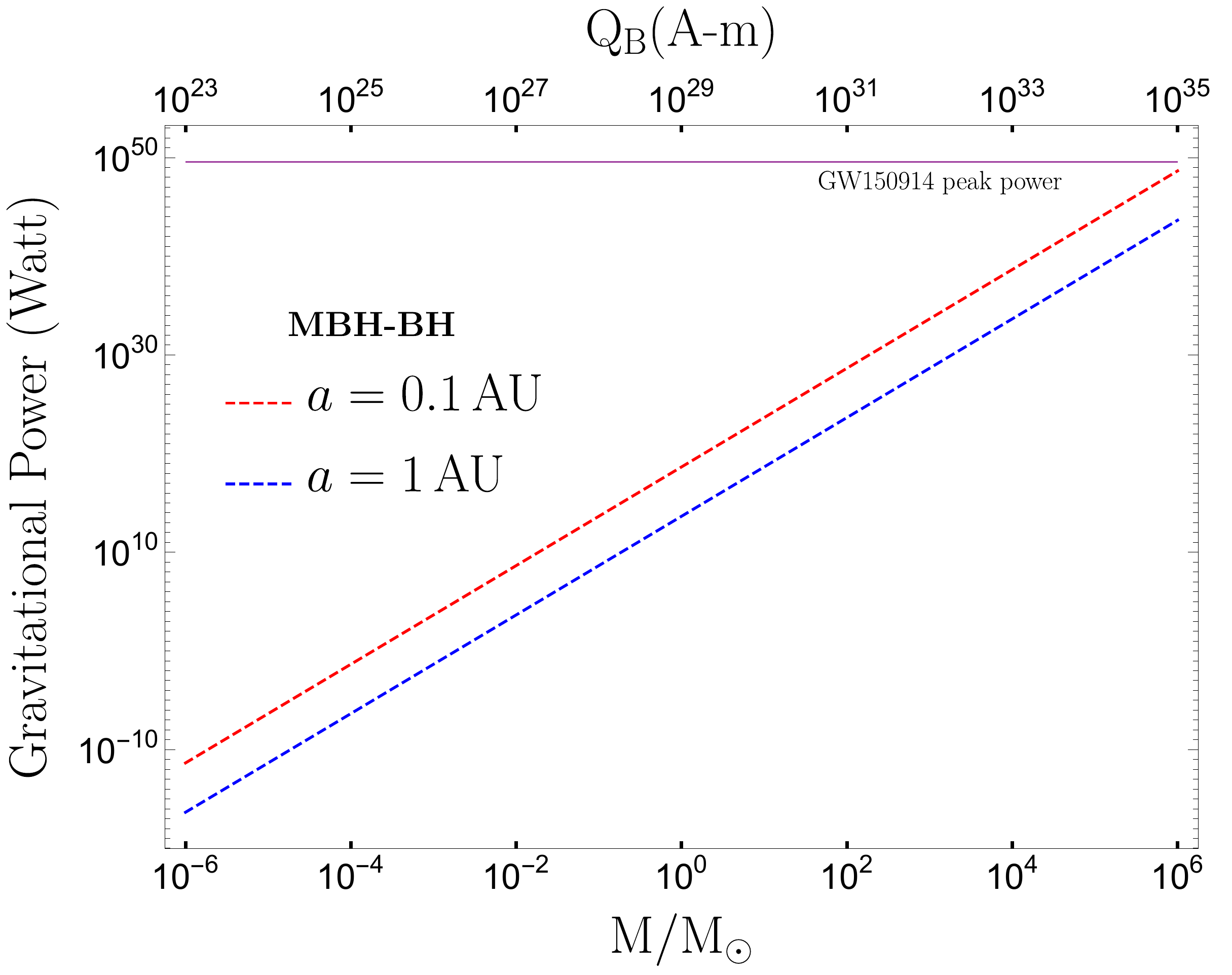}\\
\includegraphics[scale=0.18]{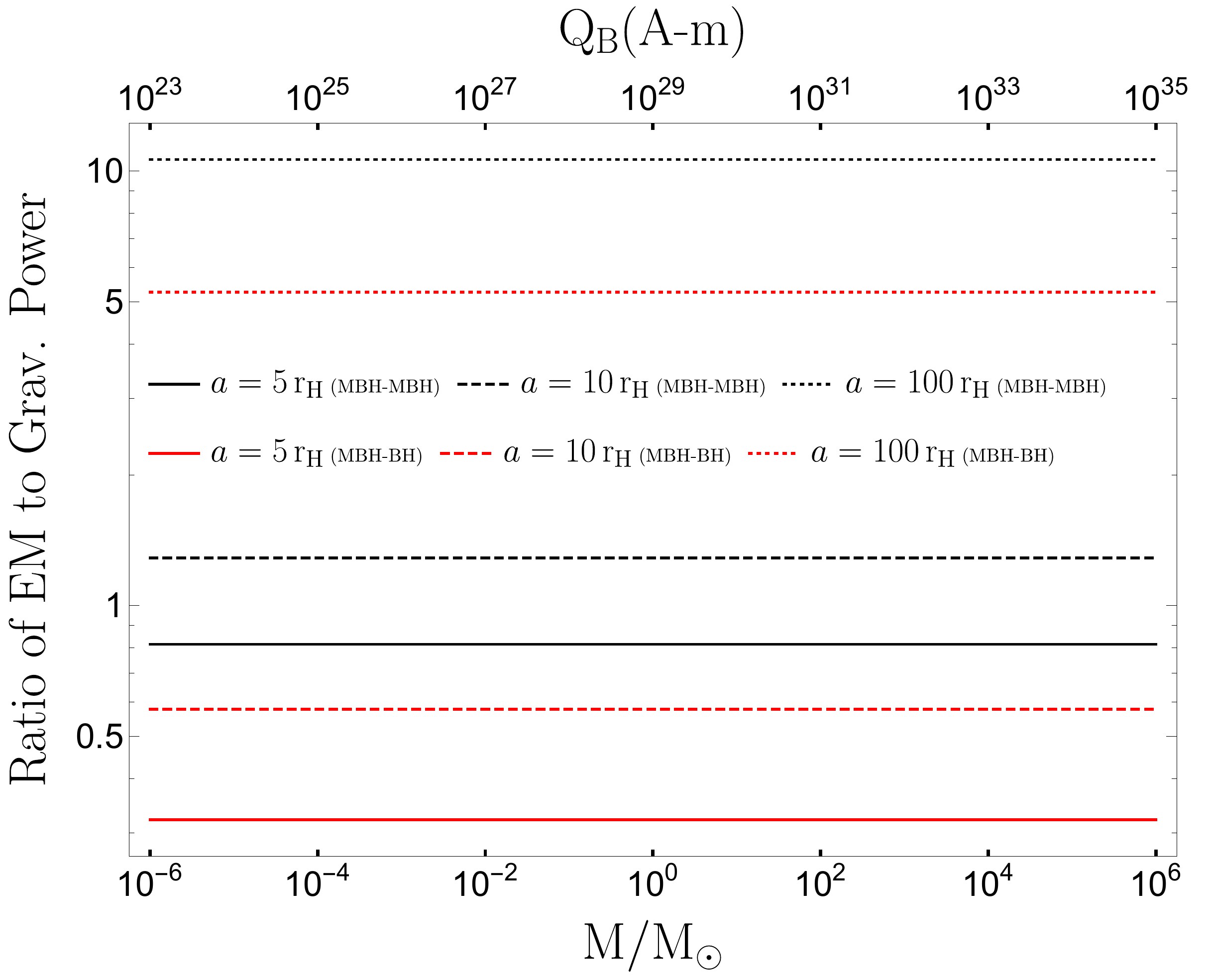} & \includegraphics[scale=0.18]{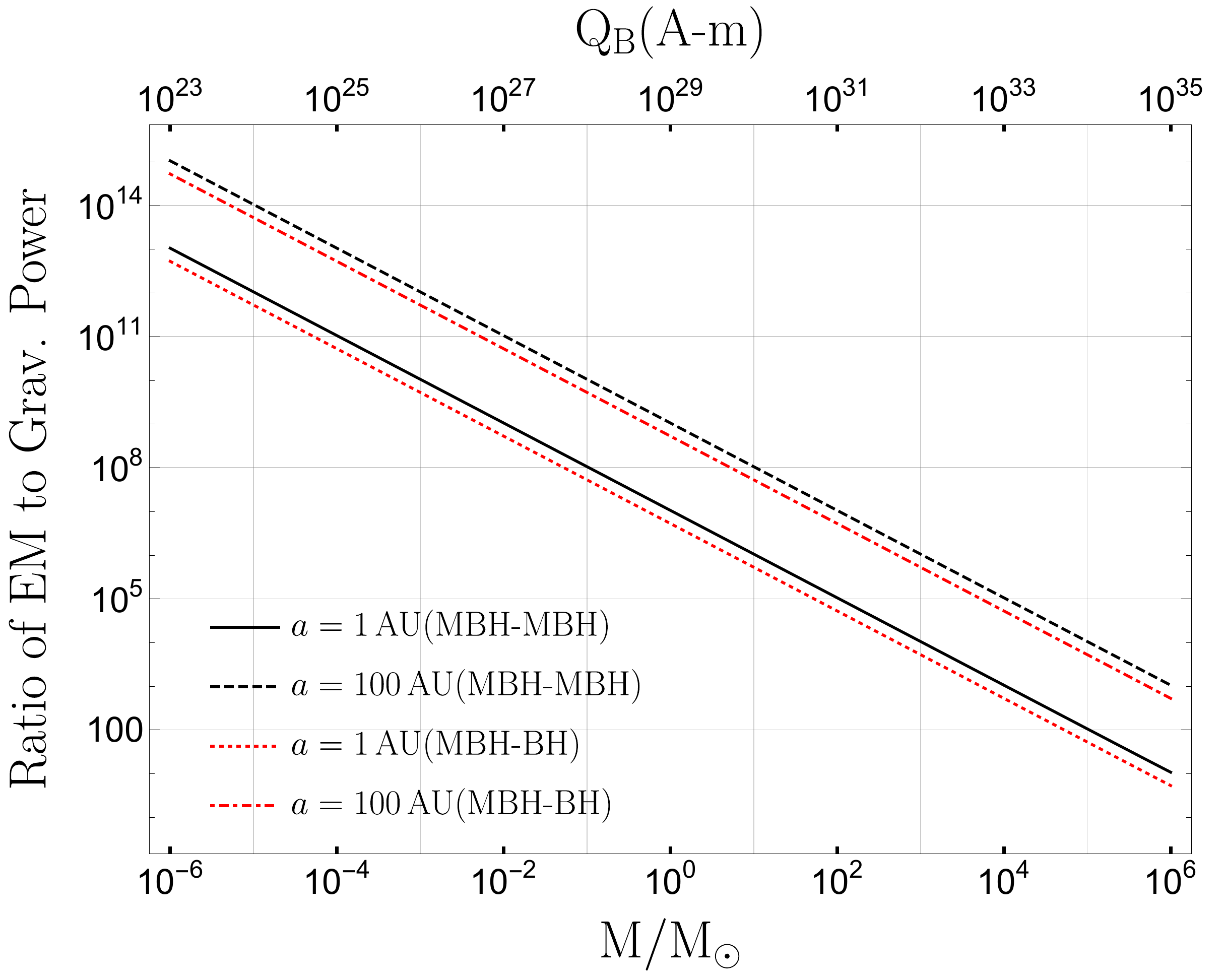}
\end{tabular}
\caption{The upper panel shows the gravitational wave power emitted by a binary blackhole system where one of them is an extremal MBH. For the MBH-MBH case, the total power radiated in gravitational waves is eight times larger than for the MBH-BH case. For comparison, the peak power emitted in gravitational waves during LIGO first-detection event GW150914\,\cite{Abbott:2016blz} is overlaid. The lower panel shows the ratio of electromagnetic power (Fig.~\ref{fig:em-power}) to the gravitational power. A guidepost may be provided by the LIGO event GW170817\,\cite{TheLIGOScientific:2017qsa}. This neutron star merger event also had a relatively weak Short Gamma Ray Burst ($\lesssim 2\,\mathrm{s}$) counterpart---GRB170817A\,\cite{Monitor:2017mdv, GBM:2017lvd}. For this case, a very crude estimate of the ratio of the isotropic equivalent electromagnetic power\,\cite{Monitor:2017mdv,GBM:2017lvd,Wang:2017rpx} to gravitational power\,\cite{TheLIGOScientific:2017qsa} gives a value $\lesssim 10^{-4}$.}
\label{fig:emtogwpow}
\end{center}
\end{figure}
%%%%%%%%%%%%%%%%%%%%%%%%%%%%%%%%%%%%

For two binary black holes in an elliptic orbit, with eccentricity $e$, the total radiated power in gravitational waves is given by\,\cite{Peters:1963ux}
\begin{equation}
P_{\text{\tiny{GW}}}^{\text{\tiny{bin.}}}=\frac{8 G_N}{5c^5} M^2 a^4 \omega^6~\frac{1}{\left( 1-e^2 \right)^{7/2}} \left( 1+\frac{73}{24} e^2+\frac{37}{96} e^4\right) \; .
\label{eq:gwpow}
\end{equation}
The binary system radiates both energy and angular-momentum through gravitational waves, and through electromagnetic emissions when charged. A careful study of the orbital evolution suggests (see for instance\,\cite{Maggiore:1900zz}) that in general the eccentricities decrease quite rapidly due to the radiation back reaction. Thus a fair approximation is to consider $e\rightarrow 0$ in Eq.\,(\ref{eq:gwpow}), and focus on quasi-circular orbits in the early stages of the inspiral.

In Fig.\,\ref{fig:emtogwpow} we show the total gravitational wave power, as well as the ratio of the electromagnetic to gravitational power emitted. Note that due to the slightly different Keplerian relations, when expressed solely as a function of the inter black hole separation, the total gravitational power radiated is different for the MBH-BH and MBH-MBH cases. The latter is about eight times larger than the former, for a fixed separation. For providing some intution, the peak gravitational wave power observed in the first LIGO detection event GW150914\,\cite{Abbott:2016blz} is also shown. The GW170817 neutron star merger event\,\cite{TheLIGOScientific:2017qsa} had a very weak electromagnetic counterpart\,\cite{Monitor:2017mdv, GBM:2017lvd}, and a rough estimate of the electromagnetic to gravitational power ratio in that case gives around $10^{-4}$ or less. Thus, in early stages of the inspiral the electromagnetic power generally exceeds the gravitational power output.

As the binary system radiates electromagnetic and gravitational waves, the radial separation between the black holes decrease. As long as $\dot{\omega}/\omega^2 \ll 1$, this evolution is well-approximated by an adiabatic sequence of quasi-circular orbits. In the regime we are making our approximations, this is satisfied to a very good extent.

 As in the last section, let us first consider the case of a neutral black hole and an MBH undergoing inspiral (MBH-BH case). Assuming a very early epoch in their binary inspiral, when they are well separated, we may write by energy conservation
\begin{equation}
\left(\frac{d \mathcal{E}}{dt}\right)^{\text{\tiny{MBH-BH}}}=-\frac{8 G_N}{5c^5} M^2 a^4 \omega^6 - \frac{\mu_0 Q_B^2}{24\pi c^3}  a^2 \omega^4 \; ,
\label{eq:freqevolvbhmbh}
\end{equation}
Here, the Keplerian relations give
\begin{eqnarray}
\mathcal{E}~&=&- \frac{G_N M^2}{2a} \; , \nn \\
\omega^2 &=& \frac{2 G_N M}{a^3} \; . 
\label{eq:keplrelbhmbh}
\end{eqnarray}
with the association $Q_B=\sqrt{\frac{\mu_0}{4 \pi  G_N}} M$, for the extremal case we are interested in.

Eq. (\ref{eq:freqevolvbhmbh}) then leads to an evolution equation for the binary system's orbital frequency
\begin{equation}
\left( \frac{\dot{\omega}}{\omega^2} \right)^{\text{\tiny{MBH-BH}}}=\frac{96}{5 c^5} \left( G_N \mathcal{M}_{\text{\tiny{ch}}}\right)^{5/3} \omega^{5/3}+\frac{\mu_0}{4\pi c^3} \left( \frac{Q_B^2}{M}\right) \omega \; .
\label{eq:orbfreqevolvbhmbh}
\end{equation}
The chirp mass as usual is defined by $ \mathcal{M}_{\text{\tiny{ch}}}=\mu^{3/5} (M_1+M_2)^{2/5}$, where $\mu=M_1 M_2/(M_1+M_2)$ is the reduced mass of the binary system. In our case, with the simplified assumption of equal masses, we therefore have $ \mathcal{M}_{\text{\tiny{ch}}}=M/\sqrt[5]{2}$. 

Since the gravitational waves from a binary black hole system is predominantly emitted at a frequency $\omega_{\text{\tiny{GW}}}=2 \omega$\,\cite{Maggiore:1900zz}, Eq. (\ref{eq:orbfreqevolvbhmbh}) may be translated into an evolution for the emitted gravitational wave frequency
\begin{eqnarray}
\left(\frac{d\nu_{\text{\tiny{GW}}}}{dt}\right)^{\text{\tiny{MBH-BH}}}&=&\frac{96\,\pi^{8/3}}{5 c^5} \left( G_N \mathcal{M}_{\text{\tiny{ch}}}\right)^{5/3} \nu^{11/3}_{\text{\tiny{GW}}}\nn \\
&+& \frac{\pi \mu_0}{4c^3} \left( \frac{Q_B^2}{M}\right) \nu^3_{\text{\tiny{GW}}} \; .
\label{eq:gwfreqevolvbhmbh}
\end{eqnarray}

The approximate adiabatic time evolution of the radial separation may also be deduced from Eqs. (\ref{eq:keplrelbhmbh}) and (\ref{eq:orbfreqevolv}) as
\begin{equation}
\left(\frac{da}{dt}\right)^{\text{\tiny{MBH-BH}}} =  -\frac{128 \sqrt[5]{8}}{5 c^5} \left( G_N \mathcal{M}_{\text{\tiny{ch}}}\right)^{3} ~a^{-3}-\frac{\mu_0 G_N}{3\pi c^3}  Q_B^2 ~a^{-2}\; . \\
\end{equation}

Consider now the case of two oppositely charged MBHs in early stages of their inspiral. From Eqs. (\ref{eq:empow}) and (\ref{eq:gwpow}) the evolution is therefore approximately given by
\begin{equation}
\left(\frac{d \mathcal{E}}{dt}\right)^{\text{\tiny{MBH-MBH}}}=-\frac{8G_N}{5c^5} M^2 a^4 \omega^6 - \frac{\mu_0 Q_B^2}{6\pi c^3}  a^2 \omega^4 \; .
\label{eq:freqevolv}
\end{equation}
Here, we now have
\begin{eqnarray}
\mathcal{E}~&=&- \frac{G_N M^2}{a} \; , \nn \\
\omega^2 &=& \frac{4 G_N M}{a^3} \; . 
\label{eq:keplrelmbhmbh}
\end{eqnarray}
For the extremal case of interest we have $Q_B=\sqrt{\frac{\mu_0}{4 \pi  G_N}} M$ again. The additional factors of two in Eq.\,(\ref{eq:keplrelmbhmbh}), relative to Eq.\,(\ref{eq:keplrelbhmbh}), are due to the fact that in the extremal case, the Coulombic attraction is exactly equal to the gravitational attraction and modifies the Keplerian relations in that manner.

%%%%%%%%%%%%%%%%%%%%%%%%%%%%%%%%%%%%
\begin{figure}[!ht!]
\begin{center}
\begin{tabular}{cc}
\includegraphics[scale=0.18]{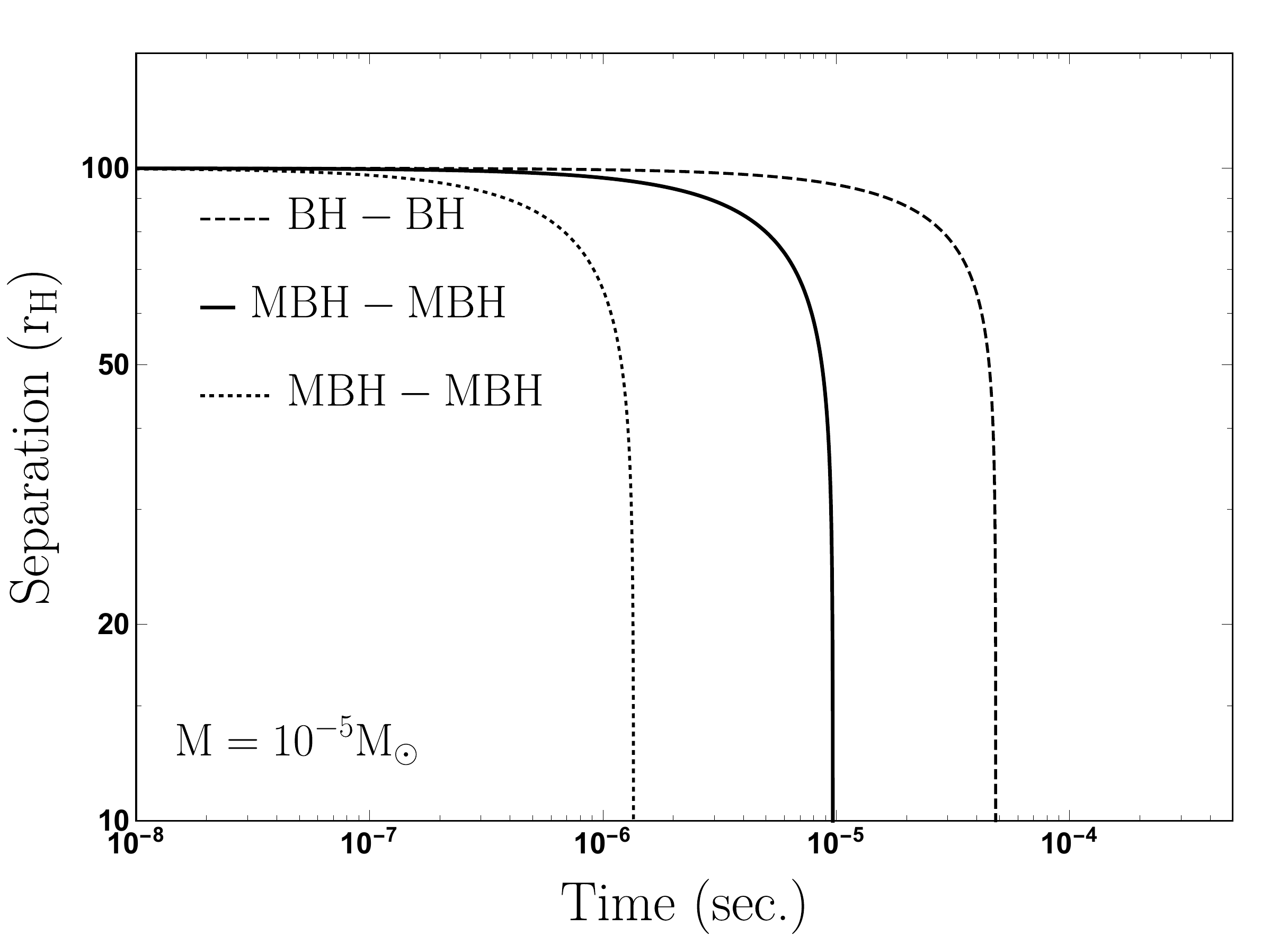}  & \includegraphics[scale=0.18]{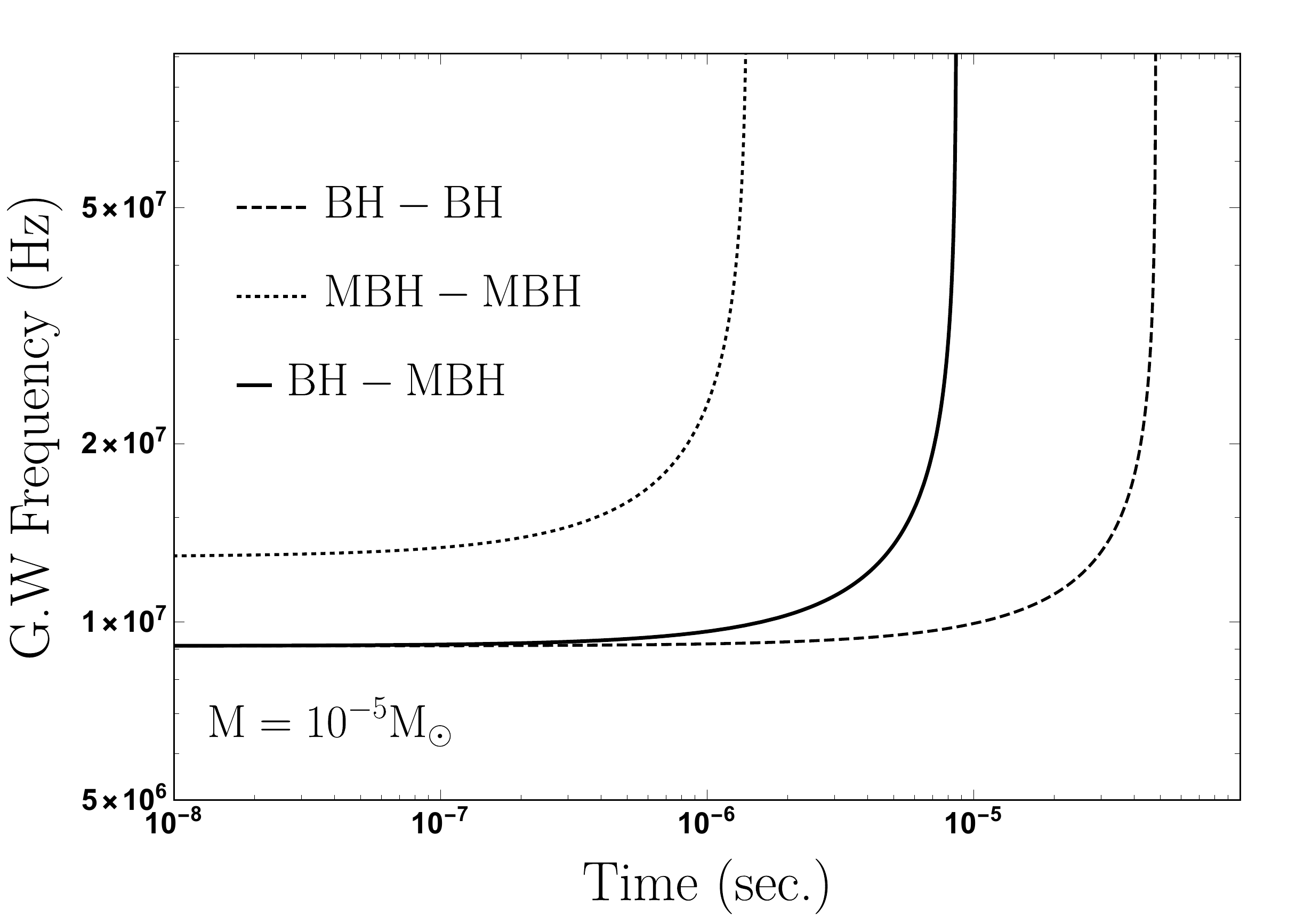} \\
\includegraphics[scale=0.18]{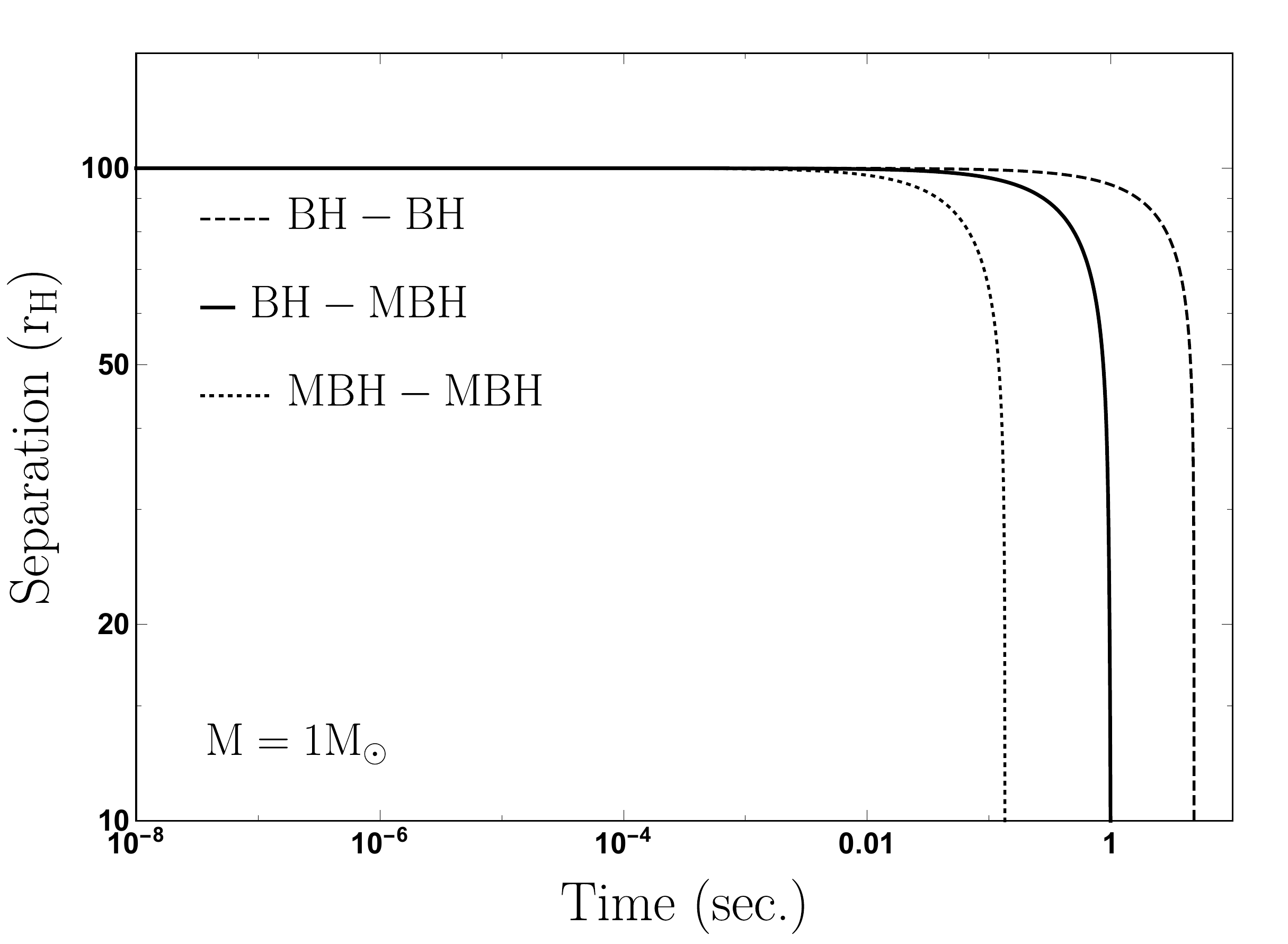} & \includegraphics[scale=0.18]{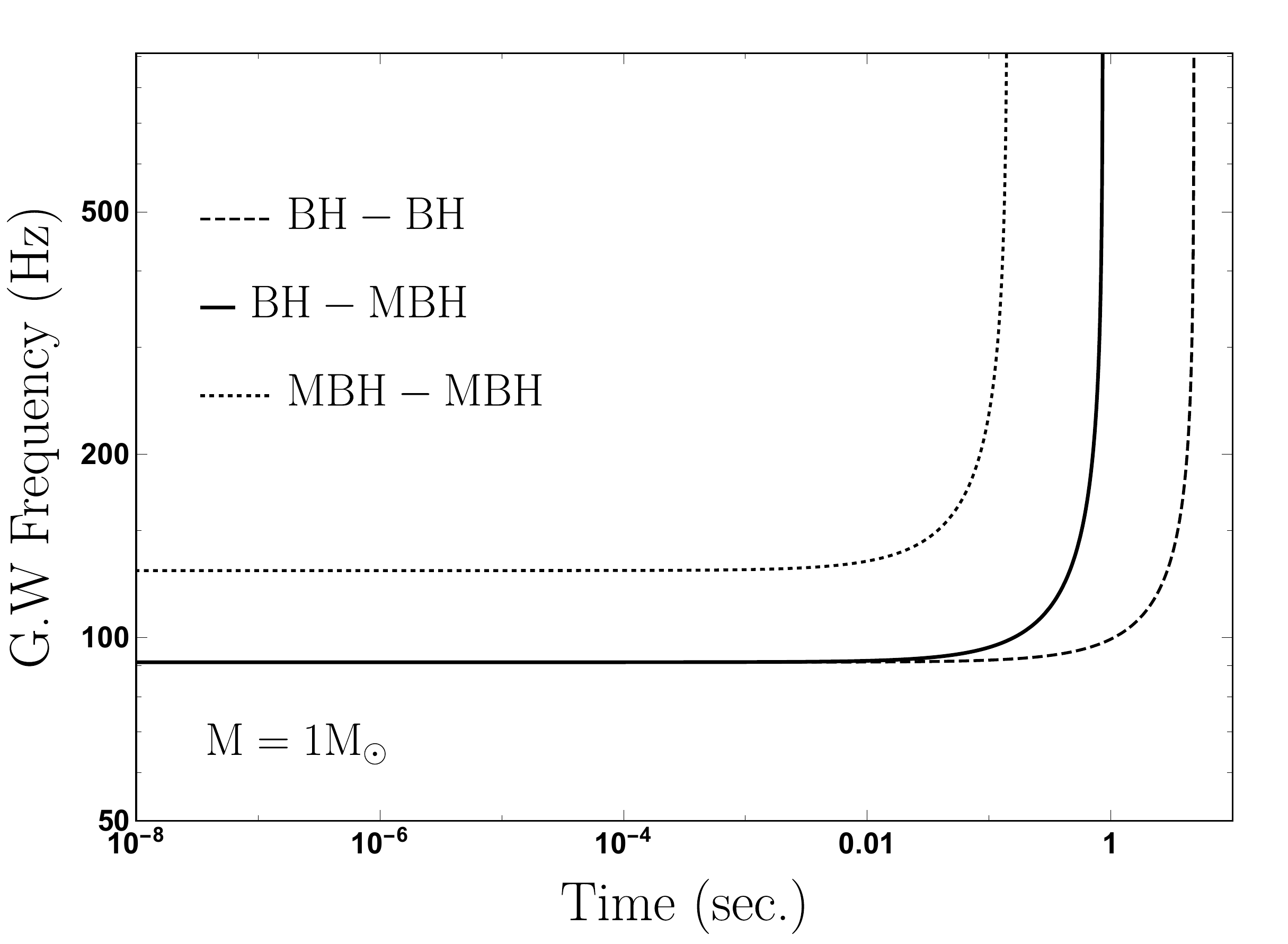} \\
\includegraphics[scale=0.18]{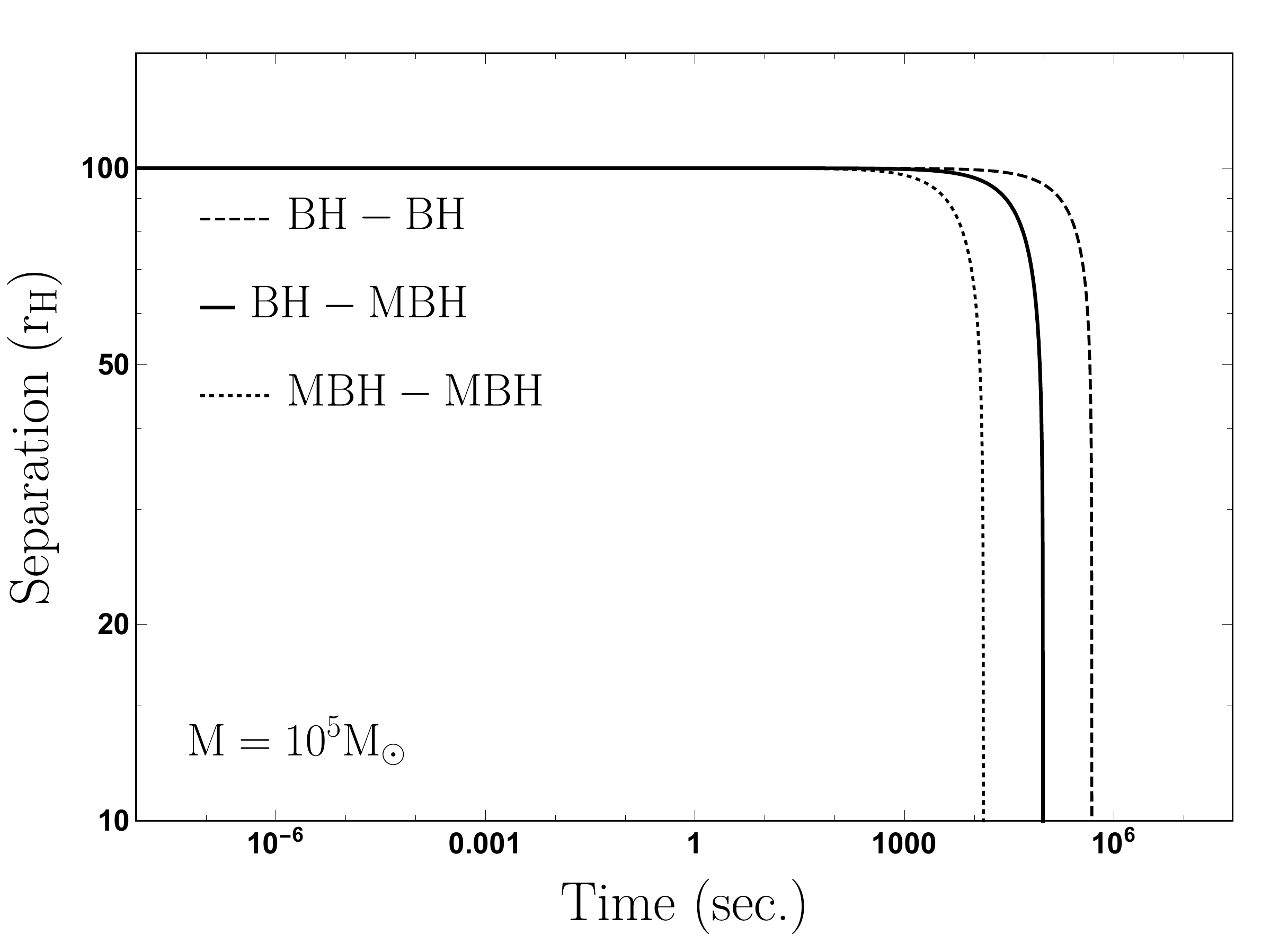} & \includegraphics[scale=0.18]{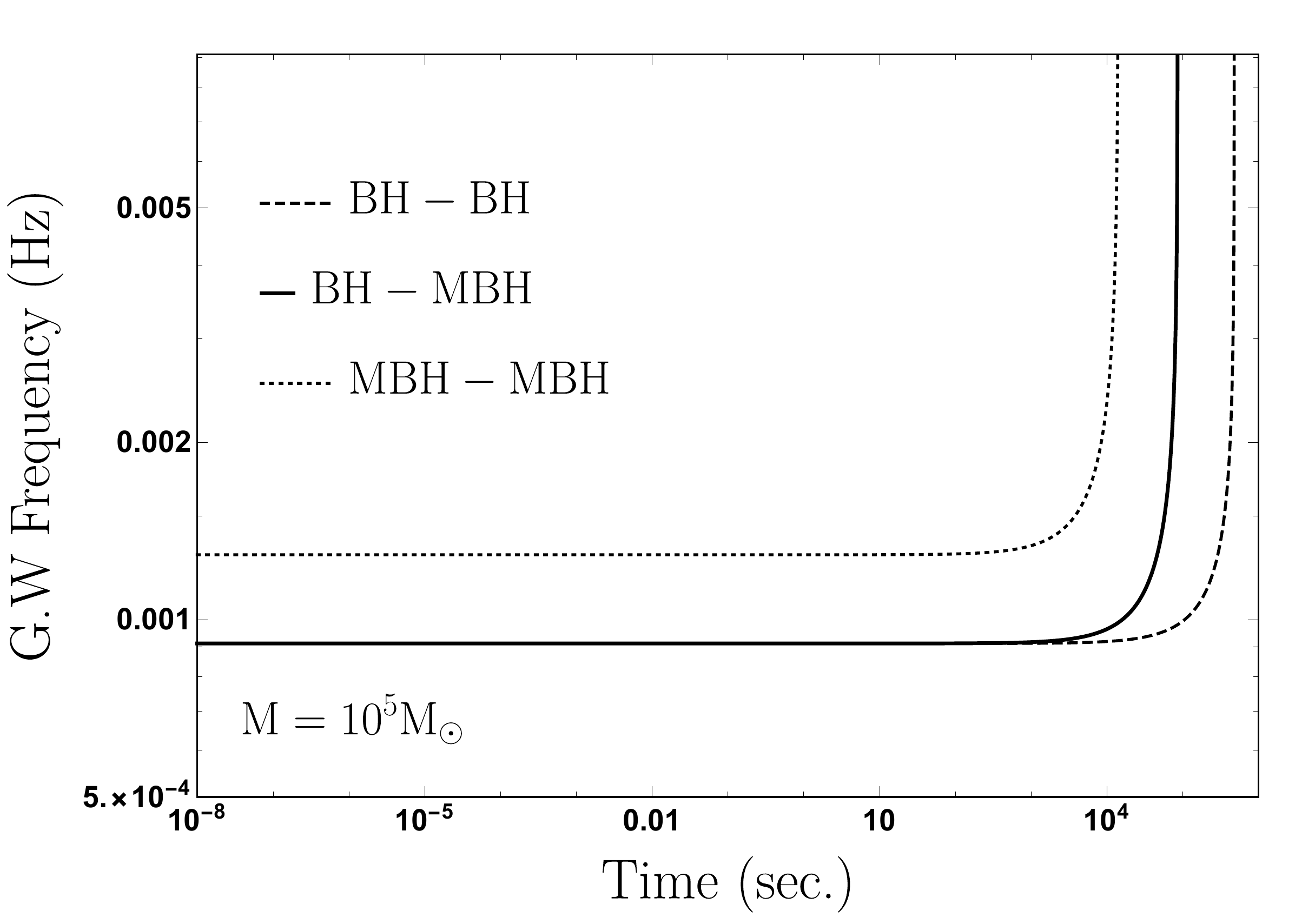} 
\end{tabular}
\caption{Evolution of the inter black hole separations and frequencies for binary black hole systems in the three cases discussed in the text. Three prototypical mass values are taken from the range of interest to us, to illustrate features and variations. One observes that the gravitational wave frequency and inter black hole separations evolve differently in the three cases of interest. As expected, the onset of chirping of the gravitational wave frequency occurs first for the MBH-MBH case.}
\label{fig:gwplots}
\end{center}
\end{figure}
%%%%%%%%%%%%%%%%%%%%%%%%%%%%%%%%%%%%

The orbital frequency of the binary system therefore evolves now as
\begin{equation}
\left(\frac{\dot{\omega}}{\omega^2}\right)^{\text{\tiny{MBH-MBH}}}=\frac{96 \sqrt[3]{4}}{5 c^5} \left( G_N \mathcal{M}_{\text{\tiny{ch}}}\right)^{5/3} \omega^{5/3}+\frac{\mu_0}{\pi c^3} \left( \frac{Q_B^2}{M}\right) \omega \; .
\label{eq:orbfreqevolv}
\end{equation}

Eq.\,(\ref{eq:orbfreqevolv}), similar to before, may be translated into an evolution equation for the emitted gravitational wave frequency
\begin{eqnarray}
\left(\frac{d\nu_{\text{\tiny{GW}}}}{dt}\right)^{\text{\tiny{MBH-MBH}}}&=& \frac{96 \sqrt[3]{4}\,\pi^{8/3}}{5 c^5} \left( G_N \mathcal{M}_{\text{\tiny{ch}}}\right)^{5/3} \nu^{11/3}_{\text{\tiny{GW}}}\nn \\
&+& \frac{\pi \mu_0}{c^3} \left( \frac{Q_B^2}{M}\right) \nu^3_{\text{\tiny{GW}}} \; .
\label{eq:gwfreqevolv}
\end{eqnarray}

Eqs. (\ref{eq:keplrelmbhmbh}) and (\ref{eq:orbfreqevolv}) now give the adiabatic evolution of the inter black hole separation as
\begin{equation}
\left(\frac{da}{dt}\right)^{\text{\tiny{MBH-MBH}}} =  -\frac{512 \sqrt[5]{8}}{5 c^5} \left( G_N \mathcal{M}_{\text{\tiny{ch}}}\right)^{3} ~a^{-3}-\frac{8\mu_0 G_N}{3\pi c^3}  Q_B^2 ~a^{-2} \; .
\end{equation}

The complete gravitational waveform has the form
\begin{eqnarray}
h_+(t)&=&\frac{G_N M a^2\omega^2}{c^4 r}   (1+\cos^2\theta) \cos (2 \omega t) \; , \nn \\
h_\times(t)&=&\frac{2 G_N M a^2\omega^2}{c^4 r}   \cos\theta  \sin (2 \omega t) \; .
\end{eqnarray}
Here, $\theta$ and $r$ are the orientation and distance to the oberver, with $\theta=\pi/2$ corresponding to the orbital plane of the inspiralling black holes. The waves are circularly polarized along the direction perpendicular to the orbital plane and linearly polarized along the orbital plane direction. We may plot the time evolution of the gravitational wave frequencies and inter black hole separation for the two cases of interest, and contrast it with the case of two neutral black holes undergoing inspiral (BH-BH case). This is illustrated in Fig.\,\ref{fig:gwplots}. The onset of chirping, of the gravitational wave frequencies, in the three cases display distinct characteristics. 

As an aside, we must mention that for precise detection of gravitational wave signals, via current matched filtering techniques, in late stages of the compact binary inspiral it is crucial to include post-Newtonian (PN) corrections\,\cite{PhysRevD.47.R4183, PhysRevD.52.821,Jaranowski:1997ky, PhysRevD.57.6168, Blanchet:2001ax,Blanchet:2006gy}. Defining
\begin{equation}
\delta=(2 G_N M\omega/c^3)^{2/3}\sim \mathcal{O}(v^2/c^2)\; ,
\end{equation}
this would modify the first terms on the right hand side of Eqs.\, (\ref{eq:orbfreqevolvbhmbh}) and (\ref{eq:orbfreqevolv}) by factors of the form 
\begin{equation}
g(\delta)=\sum_{n=0}^{N} c_{n/2}^{\text{\tiny{PN}}} ~ \delta^{n/2}\; .
\end{equation}
The coefficients $ c_{n/2}^{\text{\tiny{PN}}}$ have been computed to $n=7$\,\cite{PhysRevD.47.R4183, PhysRevD.52.821,Jaranowski:1997ky, PhysRevD.57.6168, Blanchet:2001ax,Blanchet:2006gy}. Some aspects of the PN corrections in the context of dyonic black holes were discussed in\,\cite{Liu:2020vsy}. Dyonic black hole inspirals have richer features compared to purely magnetic (or electric) black holes. For instance, as pointed out and analysed in\,\cite{Liu:2020vsy}, in the PN limit, there may be non-central, angular momentum dependent forces, and other contributions, that cause the binary orbits to display complex trajectories.

 %%%%%%%%%%%%%%%%%% Section %%%%%%%%%%%%%%%%%%%%%%%%%%%
\section{Summary and conclusions}{
\label{sec:summary}

Magnetically charged black holes\,\cite{Lee:1991vy,Lee:1991qs,Lee:1994sk,Maldacena:2020skw}, if they exist, provide a hitherto unexplored avenue to probe findamental physics and exotic quantum field theoretic phenomena in the Standard Model of particle physics, and beyond. Unlike electrically charged black holes, magnetically charged black holes may be relatively long lived and hence are promising candidates for future observations. They generically have large magnetic fields in the near horizon neighbourhoods---in many cases, approaching fields many of orders of magnitude larger than neutron stars like Magnetars\,\cite{Mereghetti:2008je}. In some regions, the field is even strong enough to provide electroweak symmetry restoration\,\cite{Maldacena:2020skw}.

We explored various possible astrophysical signatures and limits on such objects in this work, in the extremal and near-extremal limits, over the full range of astrophysically interesting mass ranges. We have added new results and estimates in this work, while also complementing recent studies\,\cite{Bai:2020spd, Liu:2020vsy} that have appeared.

We pointed out that horizons and innermost stable circular orbits in the vicinity of extremal MBHs, which set a characteristic  distance for various astrophysical phenomena such as accretion and the beginning of plunge phase during binary inspirals, show a distinct imprint of their intrinsic nature. In the case of a positive cosmological constant, there is also an intriguing qualitative feature embodied by the appearance of an outer stable circular orbit. 

Current measurements of galactic magnetic fields and dark matter density measurements were found to furnish interesting bounds on these objects across the full mass range of astrophysical relevance. In the lower mass ranges, this include cases where an electroweak corona forms near the horizon, and electroweak symmetry is restored in macroscopically large regions. Considering galactic magnetic field regeneration times, we were able to place strong bounds on magnetic black holes as dark matter candidates, with the conclusion that they are unlikely to constitute a significant component of dark matter in our universe. Within astrophysical and modelling uncertainties, our results are consistent with the seminal result found in\,\cite{Bai:2020spd}. 

Even for other masses, we pointed out that the fields are still generally enormous---enough to furnish spectacular electromagnetic astrophysical signatures, as well as affect gravitational waveforms during binary inspirals. Considering binary inspirals of MBH-BH, as well as MBH-MBH, we noted that the electromagnetic fluxes are extremely large with very characteristic angular profiles, with respect to the orbital plane. The total electromagnetic power emitted we found may even overwhelm some of the most energetic Gamma Ray Bursts known to date\,\cite{Abdo:2009zza}. In this context, we speculated on connections that MBHs may have to mysterious phenomena in the universe, like Fast Radio Bursts\,\cite{Lorimer:2007qn,Thornton:2013iua}. Interestingly, in the extremal MBH case, for a fixed separation quantified as a multiple of the horizon length, the electromagnetic power radiated is independent of the MBH mass. 

We analysed some aspects of the gravitational wave emissions from binary systems, in both the MBH-BH and MBH-MBH cases, and contrasted it with the conventional case when both black holes are neutral. For a fixed inter black hole separation, the total gravitational wave power emitted by the MBH-MBH system is larger than the MBH-BH system, owing to the different factors in the Keplerian relations. We also find that the electromagnetic emitted power dominates the gravitational wave emitted power in the early inspiral phase. The gravitational waveform for MBH-BH and MBH-MBH display interesting time evolutions, and were observed to have distinct evolution histories compared to BH-BH binary inspirals. The onset of the frequency chirping occurs much before for the MBH-MBH and MBH-BH cases. The binary merger time-scale, as inferred from the evolution of the inter black hole separation, is quicker for these cases, as expected.

A deeper exploration of some of these and other astrophysical pointers is left for future work. As already mentioned earlier, an interesting facet to ponder on is how a magnetic charge affects the ringdown and quasinormal modes in the post-merger phase\,\cite{Mellor:1989ac, Andersson:1996xw,Kokkotas:1999bd,Natario:2004jd}. An understanding of these aspects\,\cite{Maggio:2020jml} is especially pertinent in the broader context of understanding the nature of dark compact objects\,\cite{Cardoso:2019rvt}. In a similar vein, further exploration of potential near-horizon phenomena and effects on black hole shadows\,\cite{Allahyari:2019jqz} may be interesting future directions to pursue. It may be hoped that some of the astrophysical signatures, whose characteristics we have broached in the present work, may pave the way to the discovery of these exotic compact objects, if they indeed exist in our universe. 

 }
 
%%%%%%%%%%%%%% Acknowledgements %%%%%%%%%%%%%%%%%%%%%%%%
\section*{Acknowledgments}
We thank Y. Bai, J. Berger, S. Jain, M. Korwar, R. Loganayagam, and J. Maldacena for discussions, and thank D. Sachdeva for help with figures. 
 D. Ghosh acknowledges support through the Ramanujan Fellowship and the MATRICS grant of the Department of Science and Technology, Government of India. A. Thalapillil would like to acknowledge support from an Early Career Research award from the Department of Science and Technology, Government of India.
 
%%%%%%%%%%%%%%%%%%%%%%%%%%%%%
\appendix

%%%%%%%%%%%%%%%%%%%%%%%%%%%%%%
\section{Hawking temperature of MBH horizons in asymptotically de Sitter spacetime}
\label{app-A}
%%%%%%%%%%%%%%%%%%%%%%%%%%%%%

The temperature of the Black hole is given by 
\begin{equation}
    T=\frac{\kappa}{2\pi} \; ,
\end{equation}
where $\kappa$ is the surface gravity given by
\begin{equation}
\kappa^{2}=-\frac{1}{2}D^{a}\xi^{b} D_a\xi_{b} \, .
\end{equation}
Here, $\xi$ is the time-like killing vector,
\begin{eqnarray}
\xi^{a} &=& (1,0,0,0)  \; ,\\
\xi_{a}  &=& (-\Delta(r),0,0,0) \; ,\\
\end{eqnarray}
with 
\begin{equation}
\Delta(r) =  (1-\frac{2M}{r}+\frac{Q_B^{2}}{r^{2}}-\frac{\Lambda r^{2}}{3})\; .
\end{equation}

Using the above, we thus get, 
\begin{align}
\kappa^{2} &=-\frac{1}{2}g^{ac}D_{c}\xi^{0} D_a\xi_{0} \; ,\\
&=-\frac{1}{2}g^{rr}(\partial_{r}\xi^{0}+\Gamma^{0}_{r0}\xi^{0})(\partial_r\xi_{0}-\Gamma^{0}_{r0}\xi_{0}) \; .
 \end{align}
    
Using $\Gamma^{0}_{r0}=\frac{1}{2}g^{00}\partial_rg_{00}=\frac{1}{2}\frac{\Delta'(r)}{\Delta(r)}$, $g^{rr}=\Delta(r)$
and $\partial_r\xi^{0}=0$, $\partial_r\xi_{0}=-\Delta'(r)$ (rest of the terms are zero either because the Christoffel symbols are zero or because
derivative of a component of the metric vanishes), we further get
\begin{align}
    \kappa^{2}=\frac{\Delta'(r)^{2}}{4}\; ,\\
    \implies\kappa=\frac{|\Delta'(r)|}{2} \; .
\end{align}

Hence, the temperature of the black hole at coordinate distance `r' is given by 
\begin{align}
T(r) &=\frac{|\Delta'(r)|}{4\pi} \; ,\\
\end{align}
with
\begin{equation}
 \Delta'(r) =\frac{2M}{r^{2}}+\frac{2Q_B^{2}}{r^{3}}-\frac{2\Lambda r}{3} \; .
\end{equation}

Using the expressions for $r_0$ and $r_c$, from the text, we get for $M=|Q_B|$
\begin{align}
          \Delta'(r_0)=2\sqrt{\frac{\Lambda}{3}\bigg{(}1-4M\sqrt{\frac{\Lambda}{3}}\bigg{)}}\; ,\\
          \Delta'(r_c)=-2\sqrt{\frac{\Lambda}{3}\bigg{(}1-4M\sqrt{\frac{\Lambda}{3}}\bigg{)}}\; .
\end{align}
Thus, the Hawking temperature at these two horizon radii are equal, with
\begin{equation}
    T(r_0)=T(r_c)=\frac{1}{2\pi}\sqrt{\frac{\Lambda}{3}\bigg{(}1-4M\sqrt{\frac{\Lambda}{3}}\bigg{)}} \, .
\end{equation}

%%%%%%%%%%%%%%%%%%%%%%%%%%%%%%
\section{Electromagnetic radiation involving MBHs}
\label{app-B}
%%%%%%%%%%%%%%%%%%%%%%%%%%%%%

Many of the results we require may be computed readily, utilising the superposition principle and electric-magnetic duality transformations (see for instance\,\cite{Jackson:100964}). Under the duality transformations, solutions for electric charges and currents may be converted to their magnetic charge and current analogues, through the mapping
\begin{equation}
B\rightarrow E/c~~,~~~E\rightarrow-B\,c~~,~~~Q_E\rightarrow -Q_B/c \;.
\label{eq:dualtrans}
\end{equation}

Li\'enard's generalization of the Larmor formula for electric charges gives
\begin{equation}
\frac{dP}{d\Omega}=\frac{Q_E^2}{16 \pi^2 \epsilon_0} \frac{\left | \hat{\mathfrak{r}} \times \left( \vec{\mathfrak{v}} \times \vec{a}\right)\right |^2}{(\hat{\mathfrak{r}}\cdot \vec{\mathfrak{v}})^5} \; ,
\label{eq:lgenlf}
\end{equation}
which on integrating over the full solid angle gives the total electromagnetic power emitted as
\begin{equation}
P=\frac{\mu_0 Q_E^2 \gamma^6}{6 \pi c} \left[ \left | \vec{a} \right |^2 - \left |\frac{\vec{v}\times \vec{a}}{c}\right |^2\right] \; .
\end{equation}
Here, for the particle trajectory $\vec{s}(t_r)$, $t_r$ denoting the retarded time, and observation point $\vec{r}$, we have defined $\vec{v}=\dot{\vec{s}}(t_r)$, $\vec{a}=\dot{\vec{v}}(t_r)$, $ \vec{\mathfrak{r}}=\vec{r}-\vec{s}(t_r)$ and $ \vec{\mathfrak{v}}=c \hat{\mathfrak{r}}-\vec{v}$. $\gamma$ as usual is the Lorentz factor.

To compute the power radiated in the MBH-BH case easily, we may choose a coordinate system ($S'$) whose origin is at the position of the MBH. The MBH instantaneous velocity and acceleration are perpendicular to each other. Assume that at some instant it has velocity along the $\hat{y}$ direction and acceleration along $-\hat{x}$. Then, using the electric-magnetic duality transformation of Eq.\,(\ref{eq:dualtrans}) and the generalised Larmor formula of Eq.\,(\ref{eq:lgenlf}), one gets
\begin{eqnarray}
\frac{dP^{\text{\tiny{MBH-BH}}} }{d\Omega}&&=\frac{\mu_0 Q_B^2 a^2 \omega^4}{64\pi^2c^3} \cdot \\
&& \frac{\left[\left( 1-\beta \sin\theta' \sin\phi' \right)^2- \left(1-\beta^2\right)\sin^2\theta' \cos^2\phi' \right]}{\left( 1-\beta  \sin\theta' \sin\phi' \right)^5} \nn \; .
\end{eqnarray} 
This on integrating over the full solid angle then gives the total emitted power as,
\begin{equation}
P^{\text{\tiny{MBH-BH}}} = \frac{\mu_0 Q_B^2 a^2 \omega^4}{24 \pi c^3}\; .
\end{equation}

For the MBH-MBH binary system, considering quasi-circular orbits, the fields may be computed by superposing the fields from two oscillating magnetic dipoles offset by $\pi/2$. For concreteness, let us assume that the binary is in the X-Y plane of a suitably chosen coordinate frame whose origin is at the centre of mass of the system. Then, one of the oscillating magnetic dipoles is in the $\hat{x}$ direction and the other along $\hat{y}$, albeit with the $\pi/2$ phase lag. The analogous problem with opposite electric charges moving in a circle is well-known from standard electrodynamics (see for example\,\cite{Landau:1982dva}), and gives for the retarded potentials
\begin{eqnarray}
V &=& -\frac{Q_E a \, \omega \sin \theta}{4 \pi \epsilon_0 r} \left[ \cos\phi \sin \omega t_r -\sin \phi \cos \omega t_r \right] \nn \; , \\
\vec{A} &=& -\frac{\mu_0 Q_E a \, \omega}{4\pi r} \left[ \sin \omega  t_r~\hat{x} - \cos \omega t_r~ \hat{y} \right] \; .
\end{eqnarray}
Here, $t_r=t-r/c$ is the retarded time.

Using $\vec{E}=-\vec{\nabla} V -\partial\vec{A}/\partial t$ and $\vec{B}=\vec{\nabla} \times \vec{A}$, along with the duality transformations of Eq.\,(\ref{eq:dualtrans}) we get the electromagnetic fields generated during the MBH-MBH binary inspiral as 
\begin{eqnarray}
\vec{B}^{\text{\tiny{MBH-MBH}}} &=& \frac{\mu_0 Q_B a \omega^2 }{4 \pi c^2 r} \Big[ \left\{  \cos \left(\omega t_r\right) \cos\phi +\sin\left(\omega t_r\right) \sin\phi\right\}~~~\nn \\
 && \cos\theta ~\hat{\theta} + \left\{ \sin\left(\omega t_r\right) \cos\phi - \cos\left(\omega t_r\right) \sin\phi \right\}~\hat{\phi} \Big] \; ,\nn \\
\vec{E}^{\text{\tiny{MBH-MBH}}} &=& c \left( \vec{B}^{\text{\tiny{MRN bin.}}} \times \hat{r}\right) \; .
\end{eqnarray}
The usual implicit assumptions have been made---that the observation is made in the radiation zone ($r \gg a$) and that of a perfect dipole ($a\ll \lambda_{\text{\tiny{EM}}}\ll r$). With the appropriate Keplerian relations, one may check that these are satisfied for the parameter space of interest to us.

Finally, the corresponding Poynting vector, giving the instantaneous power radiated per unit area, may then be computed for the MBH-MBH case as
\begin{eqnarray}
\vec{S}^{\text{\tiny{MBH-MBH}}} &=& \frac{1}{\mu_0} \left( \vec{E}^{\text{\tiny{MBH-MBH}}}  \times \vec{B}^{\text{\tiny{MBH-MBH}}}  \right)\; , \nn \\
&=& \frac{\mu_0 Q_B^2 a^2 \omega^4}{16 \pi^2 c^3 r^2} \left[ 1-\left\{ \sin\theta \cos\left(\omega t_r -\phi \right)\right\}^2\right]~\hat{r} \; .~~~~~~
\end{eqnarray}

%%%%%%%%%%%%%%%%%%%% Bibliography %%%%%%%%%%%%%%%%%%%%%%
\bibliography{MBH} 

\end{document}